\documentclass[12pt]{article}

\usepackage{geometry}
\usepackage{amsmath,amssymb}
\usepackage{graphicx}
\usepackage{booktabs}
\usepackage{tabulary} 
\usepackage{array}    
\usepackage{natbib}   
\usepackage{hyperref}
\usepackage{multirow}
\usepackage[table]{xcolor} 
\definecolor{MyLightBlue}{RGB}{220, 240, 255}
\definecolor{MySoftYellow}{HTML}{FFF2CC}
\definecolor{MySoftGreen}{RGB}{226, 239, 218}
\usepackage[english]{babel}
\usepackage[autostyle=true]{csquotes}
\usepackage[T1]{fontenc}
\usepackage{lmodern}
\usepackage{tcolorbox}
\tcbuselibrary{listings,breakable,skins}
\newtcolorbox{promptbox}{
  breakable,
  enhanced,
  colback=gray!5,
  colframe=black,
  title=LLM Prompt,
  fonttitle=\bfseries,
  verbatim,
  left=1mm,
  right=1mm,
  top=1mm,
  bottom=1mm
}
\usepackage{pifont} 
\usepackage{xcolor}
\newcommand{\cmark}{\textcolor{green!60!black}{\ding{51}}} 
\newcommand{\xmark}{\textcolor{red}{\ding{55}}}

\title{Machine Spirits: Speculation and Adaptation of LLM Agents in Asset Markets}
\author{
  \small
  \begin{tabular}{p{\textwidth}}
    \centering
    Maxime Saxena$^{1}$, Marco Pangallo$^{2}$, Cars Hommes$^{3,4}$, Fabio Caccioli$^{1,5}$, \mbox{and R. Maria del Rio-Chanona$^{1,6,7,8,\dagger}$} \\[1em]
    \small $^{1}$Department of Computer Science, University College London, UK \\
    \small $^{2}$ESOMAS Department, University of Turin, Italy \\
    \small $^{3}$ Canadian Economic Analysis Department, Bank of Canada, Canada \\
    \small $^{4}$ Faculty of Economics and Business, University of Amsterdam, Netherlands \\
    \small $^{5}$ Systemic Risk Centre, London School of Economics and Political Science, UK\\
    \small $^{6}$Complexity Science Hub, Vienna, Austria \\
    \small $^{7}$Bennett Institute for Public Policy, University of Cambridge, UK \\
    \small $^{8}$UCL Centre for Artificial Intelligence, University College London, UK 
  \end{tabular}
}

\date{\today \vspace{-2em}}

\begin{document}
\maketitle

\renewcommand{\thefootnote}{$\dagger$} 
\footnotetext{Correspondence: m.delriochanona@ucl.ac.uk}
\renewcommand{\thefootnote}{\arabic{footnote}} 

\begin{abstract}
As Large Language Models (LLMs) become increasingly integrated into financial systems, understanding their behavioural properties is crucial. Do LLMs conform to the rational expectations paradigm, do they exhibit human-like “animal spirits”, or do they instead manifest distinct \enquote{machine spirits}? We investigate these questions with a simulated financial market, exploring the behaviour of 15 LLMs spanning a range of sizes, capabilities, and providers. Our results show that LLMs exhibit a spectrum of economic behaviours, from stable coordination on the fundamental value to human-like speculative bubbles. These behaviours are generally inconsistent with the rational expectations hypothesis. We also consider an ecology of heterogeneous agents, a more realistic setting compared to markets with identical LLM agents. These mixed markets can produce outcomes which vary substantially across repeated simulations. Even the most advanced models fail to consistently stabilise the market, with price bubbles sometimes forming despite only a minority of agents naturally forming bubbles. Instead, advanced models in mixed markets adapt their forecasting strategies to the behaviour of other agents. This adaptation can allow them to successfully exploit less sophisticated counterparts and achieve higher profits, but can also contribute to increased market volatility. These findings suggest that the introduction of AI agents into financial markets fundamentally reshapes their ecology. In particular, heterogeneous populations of LLMs can generate endogenous instability, while individual-level adaptation may amplify, rather than mitigate, market volatility.
\end{abstract}

\section{Introduction}

As Large Language Models (LLMs) are increasingly integrated into financial and economic systems, both as advisory tools \citep{lloyds2025consumerdigitalindex} and as autonomous decision systems \citep{zhao2025alphaagents,nof1_techpost1_2025}, understanding their financial behaviour becomes critical. The issue is not simply whether LLMs behave well in isolation, but what happens to market dynamics and stability when a new class of agent, one that is neither a human trader nor a hand-coded algorithm, begins to participate at scale. Do these agents simply value an asset at its fundamental price, do they reproduce the bounded and speculative dynamics associated with human \enquote{animal spirits}, or do they exhibit distinct behavioural patterns of their own, what we call \textit{machine spirits}?\footnote{
We use \textit{machine spirits} to denote systematic deviations from rational expectations driven not by emotion or experience, as in humans, but by the architectures, training data, and reasoning constraints of the models themselves.}

Initial work on LLM economic behaviour focused mainly on isolated, one-shot settings, using canonical games such as the Ultimatum Game, Prisoner’s Dilemma, and Dictator Game \citep{filippas2024large, kitadai2023toward, mei2024turing, brookins2023playing}. More recent studies have moved to dynamic multi-agent environments in which behaviour unfolds over time \citep{ashery2025emergent}. In the financial domain, \citet{chen2023put} simulate auction markets, while \citet{del2025can} study feedback-driven markets and show that earlier models such as GPT-3.5 and GPT-4 can distinguish between positive- and negative-feedback regimes, but exhibit limited behavioural heterogeneity. Similarly, \citet{henning2025llm} and \citet{muntoni2025ai} find that older models show a reduced tendency to generate bubbles compared to humans, while \citet{wang2025large} report that more recent reasoning models behave as fundamentalists. These last experiments suggest that LLMs will dampen market fluctuations, particularly relative to humans and in settings without external shocks. 

This stabilising picture may reflect the limited diversity of agents in existing experiments rather than a general property of AI-populated markets. In practice, financial systems will incorporate LLMs from different providers, alongside human participants whose behaviour LLMs may amplify or counteract. A large literature in behavioural finance and market ecology shows that instability can emerge endogenously from strategic heterogeneity (e.g. \citep{scheinkman2014speculation, bouchaud2025self}). Experimental asset markets demonstrate that human participants generate price bubbles; these bubbles cannot be explained by the Rational Expectations Hypothesis, the benchmark modelling assumption that agents form unbiased, model-consistent forecasts of future prices \citep{muth1961rational, sargent1987rational}.  Instead, they are better described by boundedly rational heuristics and models such as Heuristic Switching Models, in which agents switch between forecasting rules based on recent performance \citep{ brock1998heterogeneous, anufriev2012evolutionary, hommes2021behavioral}. The market ecology literature models markets as populations of interacting strategies whose coexistence, not representative behaviour of any single strategy type, gives rise to fragility and endogenous instability \citep{may2008ecology, scholl2021market}. If different AI agents interact and adapt strategically to one another, the entry of a new class of agent need not dampen volatility; it may reshape the market in ways that generate new forms of instability. Understanding how AI affects market stability therefore requires studying not individual LLMs but heterogeneous populations of agents, their interactions, and their adaptive responses.

We approach these questions through a deliberately simple laboratory setting: a canonical asset-pricing experiment with six agents and fifty time periods, recreating the laboratory asset markets of \citet{hommes2008expectations}. Within this setting, we study a wide range of LLMs, compare their behaviour to human experimental results, examine mixed-model markets, and analyse how AI agents adapt to one another. 

Our findings reveal that different LLMs exhibit a wide range of behaviours and are driven by distinct machine spirits.  We find low-capability LLMs that do not form bubbles but are not well described by the rational expectations hypothesis, in keeping with \citet{muntoni2025ai}, and frontier models that coordinate on the fundamental value, in keeping with \citet{wang2025large}. However, we also uncover new behaviour, such as LLMs that form speculative price bubbles, more closely resembling human experimental outcomes.

In mixed-agent markets, a setting that is under-explored in the existing literature, aggregate dynamics can be highly variable: identical experimental setups can produce outcomes ranging from stable coordination on the fundamental value to repeated speculative bubbles. These bubbles arise endogenously, without external shocks. We also show that LLMs adapt strategically. Frontier models shift from fundamentalist to trend-following behaviour when interacting with trend-following agents, resembling Heuristic Switching Models \citep{brock1998heterogeneous, anufriev2012evolutionary}. Adaptation mechanisms differ across models; for example, Gemini-3-Flash appears to employ theory-of-mind-like reasoning to increase profits, at the cost of increased market volatility.

The diversity of machine spirits has implications for economic modelling: we show that markets of LLM agents cannot generally be modelled with the rational expectations hypothesis.  The variation in macro behaviour from mixed LLM markets, not easily captured by a single average agent, also presents modelling challenges. Within the breadth of LLM behaviours, some can reproduce human results more faithfully than rational expectations; these LLMs may be better human surrogates than rational expectations in economic models. More broadly, our findings raise concerns for market stability. Mixed markets of LLM agents can generate highly variable macro-dynamics without external shocks, and agents that successfully optimise for individual profit can exacerbate market volatility in the process. 

The remainder of this paper is organised as follows. We describe the experimental design in Section \ref{sec:methods}, then present our main results in Section \ref{sec:results}: the spectrum of behaviours across 15 LLMs in homogeneous markets, the variable macro dynamics of mixed-agent markets, and the strategic adaptation of frontier models among trend-followers and how this adaptation affects volatility. We examine data leakage in Section \ref{subsec:dataleakage} and discuss implications and limitations in Section \ref{sec:conclusion}. The appendices provide robustness checks, varying temperature, memory, and bubble definitions across multiple measures, alongside associational and causal analyses of what drives bubble formation.

\section{Methods}
\label{sec:methods}

\subsection{A laboratory asset market with positive feedback}
\label{subsec:experimental design}
Our basis is the market experiment with human participants done by \citet{hommes2008expectations}.  The experiment simulates an artificial asset market with two key components: a fundamental price determined by the asset's economic fundamentals, and a realised market price that depends on participants' expectations. Six participants interact in each market over 50 periods. The realised price at time $t$, $p_t$, is determined by a market equilibrium equation
\begin{equation}
\label{eq:realisedprice}
p_t = \frac{1}{1+r}\left(\frac{1}{6}\sum_{h=1}^{6} p^e_{h,t+1} + \bar{y}\right),
\end{equation}
where $p^e_{h,t+1}$ represents participant $h$'s price prediction for period $t+1$, $r$ is the risk-free interest rate (5\% per period), and $\bar{y}$ is the mean dividend (3 guilders per period). This market equilibrium equation creates a positive feedback market: when participants predict higher future prices, the realised current price increases accordingly. To prevent prices from growing without bound, predictions are capped at a maximum of 1000, though participants only discover this constraint if they attempt to exceed it.

Each participant plays the role of a financial advisor to a pension fund and at time $t$ must predict the risky asset's price for time $t+1$ (participants at time $t$ only know realised prices up to time $t-1$ when making their predictions, making this a two time step prediction). Participants are incentivised through earnings that depend inversely on their prediction errors. The reward participant $h$ earns in period $t$ is
\begin{equation}
\label{eq:earnings}
e_{ht} = \max \left\{ 1300 - \frac{1300}{49}(p_t - p_{ht}^e)^2,0 \right\},
\end{equation}
where earnings decrease quadratically with forecast error and reach zero when errors exceed 7. Participants are told their predictions influence their pension fund's demand for the asset (higher predictions lead to greater demand), and that market prices result from the equilibrium between aggregate demand and fixed supply. 

At time $t$, participants have access to all realised prices up to period $t-1$, their own predictions up to period $t$,  their earnings from period $t-1$, and  their cumulative earnings. They also know the market fundamentals: the mean dividend ($\bar{y} = 3$) and risk-free rate ($r = 0.05$).  The initial price predictions (periods 1 and 2) are made with the guidance that prices will likely be between 0 and 100. However, participants do not know the exact market equilibrium equation, how many other participants are in the market, or other participants' predictions,  strategies and earnings. Participants have no communication with each other throughout the experiment.

The market has a theoretical fundamental price of $p^f = \bar{y}/r = 60$, representing the present value of expected future dividends. Since market participants know the mean dividend and risk-free rate, they should in theory be able to calculate the fundamental price. Even if participants do not successfully compute the fundamental price, as prices rise the effective dividend rate decreases ($\frac{3}{p_t}$) since $\bar{y}$ is fixed at 3. This means that the interest rate discounting has a greater impact on the price calculation and prices feel a \enquote{pull} back towards this fundamental value. The bar to create a bubble is therefore quite high, as \citet{hommes2008expectations} show that the fundamental value is locally stable, even under several boundedly rational prediction strategies. 

\subsubsection{The rational expectations hypothesis}

The rational expectations hypothesis assumes that agents form unbiased expectations for future prices that are equal to the mathematical model expectation of future prices \citep{muth1961rational}; it therefore assumes that participants know the asset pricing model (Equation \ref{eq:realisedprice}). Further, all participants are taken to know, and coordinate on, each other's expectations. 

Under these assumptions, the constant fundamental price (60) is the first solution to Equation \ref{eq:realisedprice}, found if agents coordinate on the \enquote{no bubble} condition (the expected present value of future prices tends to zero as time goes to infinity).  However, rational expectations also allows the existence of rational bubbles, where prices grow exponentially at the risk-free rate of return: $p_t = p^f + cR^t$, where $R = 1 + r = 1.05$ and $c$ is any positive constant \citep{hommes2008expectations}. Rational bubbles satisfy rational expectations because if all agents expect and coordinate on this growth rate, the capital appreciation of the risky asset offsets the lowering dividend yield, which makes holding the risky asset as appealing as holding cash (which accumulates at the risk-free rate).  During a rational bubble, when the 1000 barrier is hit and the agents learn of the cap, the rational expectations solution jumps to $p^f = 60$ as the \enquote{no bubble} condition is now mechanically fulfilled. See Appendix \ref{app:rehsolution} for a derivation of the rational expectations hypothesis solution.

\subsection{From human participants to LLM agents}

Our experiments replace human participants with LLM agents. We use a combination of locally run LLMs with open weights \citep{wolf2020transformers}, the OpenAI API and the Gemini API. Each time step occurs sequentially, with all 6 LLM agents being prompted to give their predictions at time $t$ for time $t+1$ before the realised price and earnings for time $t$ are calculated and the next time step occurs. 

\subsubsection{LLM agent settings}

We detail some of the key parameters in the translation to LLMs, how LLMs were queried and the settings that were used in our experiments.

\textbf{Model:} at the core of an LLM agent is a model such as GPT-5 Mini or Qwen3-14B. We consider a wide range of models, with sizes ranging from 7 billion parameters up to the extremely large models, such as those from OpenAI where the exact parameter counts are undisclosed (likely  200 billion or more \citep{EpochAIModels2025, epoch2024frontierlanguagemodelshavebecomemuchsmaller}). We also consider a range of five LLM families, some of which are open-weight (Qwen, OLMO, R1 and Gemma). The models we consider in this paper are listed in Table \ref{tab:models}, along with the names we use to refer to them from here on. Throughout this paper we use the medium reasoning setting where models have variable reasoning efforts. 

\begin{table}[htbp]
\centering
\renewcommand{\arraystretch}{0.95}
\begin{tabular}{lll}
\hline
\textbf{Provider} & \textbf{LLM} & \textbf{Paper name} \\
\hline
\rowcolor{gray!15} Alibaba & Qwen3-32B & Qwen3-32B \\
\rowcolor{gray!15} & Qwen3-14B & Qwen3-14B \\
\rowcolor{gray!15} & Qwen2.5-7B-Instruct & Qwen2.5 \\
Ai2 & OLMO3-7B-Think & OLMO-Think \\
& OLMO3-7B-Instruct & OLMO-Instruct \\
\rowcolor{gray!15} Deepseek & R1-Distill-Llama-8B & R1-Llama \\
\rowcolor{gray!15} & R1-Distill-Qwen-14B & R1-Qwen \\
Google & Gemini-3-Flash-Preview & Gemini-3-Flash \\
& Gemini-2.5-Flash-Preview & Gemini-2.5-Flash \\
& Gemma3-27B & Gemma \\
\rowcolor{gray!15} OpenAI & GPT-4.1 & GPT-4.1 \\
\rowcolor{gray!15} & GPT-4o Mini & GPT-4o Mini \\
\rowcolor{gray!15} & o3-mini & o3-mini \\
\rowcolor{gray!15} & o3 & o3 \\
\rowcolor{gray!15} & GPT-5 Mini & GPT-5 Mini \\
\hline
\end{tabular}
\caption{\textbf{Large language models used in this paper.} Where the short name used throughout the text differs from the full model name, it is provided in the \enquote{Paper name} column. We use a wide variety of models across different sizes, capabilities and providers.}
\label{tab:models}
\end{table}

\textbf{Temperature:} the temperature parameter of an LLM controls how random the model's response is. Lower values, near 0, are closer to deterministic outputs, while higher numbers, typically near 1, are used for more randomness and diversity of output. We use a temperature of 1 throughout this paper. This is to ensure consistency across models as some GPT models only accept this value of the parameter. We show that our results are robust to this setting in Appendices \ref{app:results:robustsinglellm} and \ref{app:results:mixedmarketstemprobust}.

\textbf{Memory:} in these experiments, we define memory as the number of an agents' own prior responses it sees as context for its next price prediction. We use a value of 2 throughout these experiments, meaning that an agent can see its last two price predictions, as well as the prior 2 user prompts.  We show that our results are robust to this parameter in Appendix \ref{app:results:robustsinglellm} and refer the interested reader to work such as in \citep{del2025can} for a more in-depth exploration.

\textbf{Seed:} this parameter is designed to allow reproducibility and encourage transparency.  LLM outputs are generally stochastic but setting the seed makes it possible to reproduce this randomness. Open-source models are all run with predetermined seeds which change across runs and which also change for every agent and every time step.  In the OpenAI API, setting the seed is said to give \enquote{(mostly) deterministic outputs across API calls}, making these results most likely replicable despite OpenAI stopping short of giving a full guarantee of reproducibility \citep{openai2026advanced}. The Google API does not allow reproducibility through the seed parameter. We use the Google API for Gemma and Gemini queries. 

\subsubsection{Prompts}

We keep instructions for the LLMs as similar as possible to those used in the human experiments. However, some minor changes are required. To interact with the LLMs in a systematic manner, we include specifications in the system prompt and each user prompt to reply in JSON format. Throughout, we design the prompts to provide clear instructions, and also to allow the LLMs space to \enquote{think} (even if they are not explicitly fine-tuned to output thinking tokens). Space for LLMs to \enquote{think} is achieved by specifying a \enquote{justification} key in the JSON output in which the LLM is asked to justify its price prediction, before outputting its prediction.  A useful side-effect of this is that we also have a logic trail for LLM agent price predictions throughout, giving us insights into why LLM agents output the prices they do. Where applicable, tabular data is formatted as a markdown table \citep{fang2024large}. 

The system prompt is identical to the explanatory text given at the start of the original experiment except for added instructions on the response format, asking that responses are in JSON format and the keys that should be used. Then, for each time step a user prompt is also created which reflects as closely as possible the information given on-screen to human participants. The only addition is a reiteration of the JSON formatting request. The original experiments instructions in \citet{hommes2008expectations} are in Appendix \ref{app:originalprompts} while details of the prompts given to LLMs are in Appendix \ref{app:LLMprompts}.

\subsubsection{Experiment flow}

As in the original experiments by \citet{hommes2008expectations}, each market has 6 agents and the experiment is run over 50 time steps throughout this paper.  At the initial time step, agents are informed that the price is likely between 0 and 100, in keeping with the original experiment.  For this first time step alone, agents are also asked to predict the first two prices, after which price predictions return to only forecasting one price. The original experiment also imposes a maximum prediction of 1000, although participants are only told about this artificial cap on predictions if they try to predict above the cap. We implement this as well, adding an extra note to the system prompt that predictions must be less than or equal to 1000 if and only if an agent attempts to predict above the cap. The sequence of each time step $t$ is therefore:

\begin{enumerate}
    \item \textbf{Instructions and context:} each LLM agent is given the experiment instructions, in the form of a system prompt, as well as their own output from the last two time steps according to the memory parameter of 2. 
    \item \textbf{Prompt:} each agent is also given a new user prompt, which includes information such as prior realised prices, their prior predictions, their earnings at the last time step and their total earnings so far. 
    \item \textbf{Prediction:} using the instructions, context and prompt, each LLM then outputs a price prediction for time step $t+1$, as well as a justification for this prediction in JSON format.
    \item \textbf{Market price and earnings calculations:} using the price predictions from all agents for time step $t+1$, the realised price for time step $t$ is calculated according to Eq. \ref{eq:realisedprice}. With the realised price for time step $t$ and the predicted prices for time step $t$ (predicted at time step $t-1$), we then calculate earnings for each agent according to Eq. \ref{eq:earnings}.
\end{enumerate}

\subsection{Quantifying mispricing, bubbles and agent dispersion}
\label{subsec:bubble_measures}

To quantify the extent of bubble formation and mispricing in experimental markets, we employ several standard measures from the experimental asset market literature: Relative Deviation (RD), Relative Deviation Maximum (RDMAX) and the Interquartile Range of prices (IQR pr). These measures capture different aspects of how prices deviate from fundamental values: the direction and magnitude of mispricing, the peak mispricing, and overall price volatility. 

We use Relative Deviation (RD) \citep{stockl2010bubble} to capture systematic overpricing or underpricing of the realised market prices relative to the fundamental value (value of 1 is a 100\% mean overpricing).  Relative Deviation Maximum (RDMAX) is defined as the peak percentage deviation from the fundamental value observed during the experiment \citep{razen2017cash, HOYER202332, jiao2025green}. This measure is particularly useful for identifying the amplitude of price bubbles. To assess overall price variability, we calculate the interquartile range of realised market prices \citep{hanaki2023forecasting}. Formal definitions, including several extra measures for robustness, are provided in Appendix \ref{app:bubblemeasures} and the results of these extra measures are given in Appendix \ref{app:results:measures}.

Bubbles are defined as a sustained overpricing where the market price remains above five times the fundamental value (300) for at least three consecutive periods, and we refer to this binary indicator as $P_{bubble}$. Given there is no universally accepted definition of experimental bubbles, a discussion of robustness, including a comparison to a bubble measure used in \citet{kopanyi2021experience, kopanyi2024role}, is found in Appendix \ref{app:robustnesschecks}. These two measures are shown to be robust across different parameter values and to agree with each other, with Cohen's kappa scores of 0.85 or above in all comparisons.

We characterise bubbles using the speed of formation and burst dynamics. When experiments produce multiple bubbles, we focus our analysis on the first bubble, as this captures the initial market dynamics before participants learn from the burst. To measure these, we first identify the bubble peak as the first local maximum within a seven-period window exceeding the threshold of 300 . The bubble start is defined as the earliest period where all subsequent prices up to the peak remain above the fundamental value.  Based on these definitions we calculate: (i) time to form ($T_{bottom \rightarrow peak}$), the number of periods from bubble start to peak; and (ii) half-life (HALF), the number of periods after the peak for prices to fall below half the peak value. These measures allow us to compare how quickly bubbles form and how rapidly they burst.

We also analyse the level of coordination and agent heterogeneity within each market. Following the methodology in \citet{hommes2008expectations}, we break down the mean agent squared error into the dispersion error and the common error according to Equation \ref{eq:agent_error}. A high dispersion error suggests that individual agent predictions are quite different from each other, while a high common error suggests that agents coordinate on an incorrect average prediction.

\begin{equation}
\label{eq:agent_error}
    \frac{1}{T_0 H} \sum_{h,t} (p_{ht}^{\text{e}} - p_t)^2 = 
        \underbrace{\frac{1}{T_0 H} \sum_{h,t} (p_{ht}^{\text{e}} - \bar{p}_t^{\text{e}})^2}_{\text{Dispersion error}} + 
        \underbrace{\frac{1}{T_0} \sum_{t} (\bar{p}_t^{\text{e}} - p_t)^2}_{\text{Common Error}}
\end{equation}

\subsection{Testing alignment with the rational expectations hypothesis}

We consider whether LLM agents behave consistently with the rational expectations hypothesis. Looking first at bubbles, we distinguish rational bubbles from speculative bubbles by testing the growth rate. For a rational bubble, the log deviation from fundamentals grows at a constant rate: if $q_t = \ln(p_t - p^f)$, then $q_{t+1} - q_t = \ln R \approx 0.049$. Following \citet{hommes2008expectations}, we calculate $q_{t+1} - q_t$ from the bubble start to the first deceleration point before the peak (where $p_t - p_{t-1} \leq p_{t-1} - p_{t-2}$). We then conduct a one-sided t-test with $H_0: q_{t+1} - q_t \leq \ln R$ against $H_1: q_{t+1} - q_t > \ln R$. Rejection of the null hypothesis indicates a speculative bubble growing faster than rational expectations would predict, consistent with boundedly rational trend extrapolation and speculation by agents.  The proportion of experiments in which a speculative bubble forms is referred to as SPEC.

The rational expectations condition from \citet{muth1961rational}, that the expectation of agents is equal to the expected value calculated by the model, can also be tested. One implication of this condition is that the expectations of each agent should be unbiased, and we follow the well-trodden path in the literature, using the method proposed in \citet{mincer1969evaluation}. We regress the realised prices on agent predictions, $p_t = \alpha_0 + \alpha_1 p_{h,t}^e + \epsilon$, and test against the null hypothesis: $\alpha_0 = 0$ and $\alpha_1 = 1$. If the null hypothesis is rejected, then we conclude that an agent forms biased predictions and their behaviour is inconsistent with the rational expectations hypothesis. The proportion of biased agents is referred to as BIAS.

\section{Results}
\label{sec:results}

\subsection{Towards a taxonomy of Machine Spirits}
\label{subsec:bounded_rationality}

\begin{figure}[htbp]
    \centering
    \includegraphics[width=\linewidth]{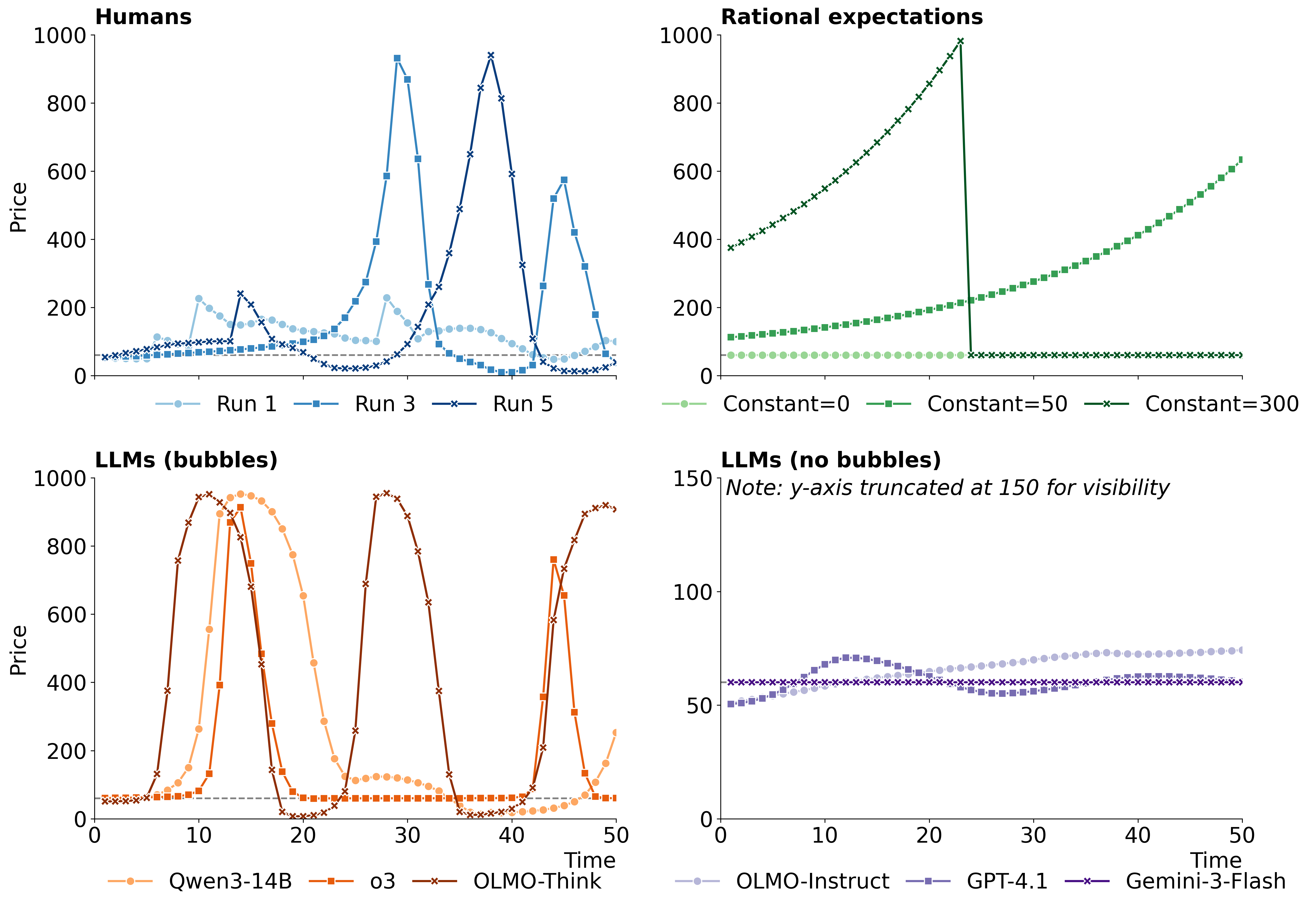}
    \caption{\textbf{LLMs give a wide range of results, from speculative bubbles to rational behaviour.} Human results are from \citet{hommes2008expectations}.  A selection of the infinite rational expectations (RE) solutions are shown, with the constants corresponding to the \enquote{c} in the RE solution $p_t = p^f + cR^t$.  See Appendix \ref{app:rehsolution} for a derivation of the RE solutions. LLM results are split into experiments in which a bubble was formed and those where there was no bubble.}
    \label{fig:intuition_plot}
\end{figure}

Just as human economic behaviour is driven by animal spirits, our results reveal a wilderness of machine spirits dictating how different LLMs act. A selection of experiments is presented in Figure \ref{fig:intuition_plot}, showing results from human participants and LLM agents as well as the rational expectations solutions. 

Several LLMs give results qualitatively similar to the bubbles formed by humans, with prices rising close to 1000 (fundamental price $p^f = 60$), before crashing down.  Other LLMs, such as OLMO-Instruct or GPT-4.1, fail to form full speculative bubbles but are not qualitatively similar to any rational expectations solutions.  Finally, some LLMs, such as Gemini-3-Flash, predict 60 throughout, making their behaviour look similar to the fundamental value solution under the rational expectations hypothesis. To further explore our results, we use the quantitative measures discussed in Section \ref{subsec:bubble_measures} and present the mean of each measure, calculated across independent experimental runs, in Table \ref{tab:bubbchar}. These measures divide the LLMs into three groups: five LLMs form bubbles, seven fail to form bubbles but are inconsistent with the rational expectations hypothesis, and three models give results similar to the rational expectations hypothesis solution predicting the fundamental value throughout.

\begin{table}[htbp]
\centering
\setlength{\tabcolsep}{3pt}
\renewcommand{\arraystretch}{0.9}
\begin{tabular}{lrrrrrrrrr}
\toprule
 &  & \multicolumn{2}{c}{Overpricing} & \multicolumn{1}{c}{Volatility} & \multicolumn{3}{c}{Bubble shape} & \multicolumn{2}{c}{Non-REH} \\
\cmidrule(lr){3-4} \cmidrule(lr){5-5} \cmidrule(lr){6-8} \cmidrule(lr){9-10} 
Model & Repeats & RD & RDMAX & IQR pr & $P_{bubble}$ & $T_{b \rightarrow p}$ & HALF & SPEC & BIAS \\
\midrule
o3-mini & 6 & 6.17 & 14.21 & 457.91 & 1.00 & 7.50 & 13.67 & 1.00 & 0.83 \\
OLMO-Think & 17 & 6.08 & 14.90 & 666.37 & 1.00 & 6.76 & 6.82 & 0.71 & 0.95 \\
\rowcolor{MySoftGreen} RE (constant = 400) &  & 3.62 & 15.28 & 504.44 & 1.00 & 16.00 & 1.00 & 0.00 & 0.00 \\
Qwen3-14B & 20 & 3.25 & 14.32 & 279.40 & 1.00 & 6.60 & 8.58 & 0.95 & 0.75 \\
\rowcolor{MySoftGreen} RE (constant = 575) &  & 2.53 & 15.61 & 0.00 & 1.00 & 9.00 & 1.00 & 0.00 & 0.00 \\
\rowcolor{MyLightBlue} Humans & 6 & 2.40 & 12.16 & 213.77 & 0.83 & 15.60 & 3.20 & 0.80 & 0.69 \\
Qwen3-32B & 15 & 2.12 & 6.13 & 228.52 & 0.40 & 12.67 & 12.33 & 1.00 & 0.40 \\
o3 & 6 & 1.42 & 12.50 & 62.50 & 1.00 & 8.17 & 2.67 & 0.17 & 0.22 \\
GPT-4o Mini & 6 & 0.26 & 0.58 & 19.38 & 0.00 & - & - & - & 1.00 \\
OLMO-Instruct & 20 & 0.14 & 0.28 & 10.51 & 0.00 & - & - & - & 0.95 \\
Qwen2.5-7B & 20 & 0.08 & 0.30 & 13.10 & 0.00 & - & - & - & 0.98 \\
R1-Qwen & 17 & 0.08 & 0.22 & 8.96 & 0.00 & - & - & - & 0.79 \\
R1-Llama & 19 & 0.02 & 0.12 & 3.70 & 0.00 & - & - & - & 0.87 \\
GPT-5 Mini & 6 & 0.01 & 0.02 & 0.50 & 0.00 & - & - & - & 0.25 \\
Gemini-2.5-Flash & 6 & -0.00 & 0.00 & 0.00 & 0.00 & - & - & - & 0.03 \\
Gemini-3-Flash & 6 & 0.00 & 0.00 & 0.00 & 0.00 & - & - & - & 0.00 \\
GPT-4.1 & 6 & 0.00 & 0.10 & 3.42 & 0.00 & - & - & - & 0.81 \\
\rowcolor{MySoftGreen} RE (constant = 0) &  & 0.00 & 0.00 & 0.00 & 0.00 & - & - & - & 0.00 \\
Gemma & 18 & -0.04 & 0.01 & 3.93 & 0.00 & - & - & - & 0.94 \\
\bottomrule
\end{tabular}
\caption{\textbf{Quantitative analysis shows a wide range of results.} Statistics are averaged across independent experimental repeats.  \enquote{RE} rows are rational expectations hypothesis solutions, and are highlighted in green. The constant in these solutions refers to the \enquote{c} in the rational bubble solution, $p_t = p^f + cR^t$. Human results from \citet{hommes2008expectations} are highlighted in blue. Relative Deviation (RD) is a measure of over-pricing measured as a multiple of the fundamental value, while RDMAX is the maximum overpricing. IQR pr is the inter-quartile range of the prices while $P_{bubble}$ is the proportion of runs which formed a bubble. The time for the bubble to form ($T_{bottom \rightarrow peak}$), half-life of the bubble (HALF), and proportion of experiments with provably speculative bubbles (SPEC) are averaged across experiments only where a bubble is formed. BIAS is the proportion of biased agents across all experimental runs.  Models are sorted by descending RD. Robustness of these results to varying temperature and memory is shown in Appendix \ref{app:results:robustsinglellm}.}
\label{tab:bubbchar}
\end{table}

Looking first at the LLMs that form bubbles, o3-mini, OLMO-Think and Qwen3-14B all consistently form bubbles which are provably speculative and driven by biased agents. However, the measures in Table \ref{tab:bubbchar} show that these three models form different shape bubbles: OLMO-Think forms the sharpest bubbles (lowest $T_{b \rightarrow p} = 6.76$ and HALF=6.82) while o3-mini takes a much longer time to come down after reaching its bubble peaks (HALF=13.67).  Meanwhile, o3 always forms a bubble, but these bubbles are not generally speculative (SPEC=0.17) and burst quickly (HALF=2.67).  Unique among the LLMs we consider is Qwen3-32B; it is the only LLM to produce variable macro behaviour, forming bubbles only 40\% of the time. When Qwen3-32B does form a bubble, the bubbles are consistent with bounded rationality.  When it does not form a bubble Qwen3-32B has some runs where it coordinates on the fundamental price, and some runs with biased agents, non-zero mispricing and volatility, not well captured by the rational expectations hypothesis. 

 Turning to the LLMs that do not form bubbles, GPT-5 Mini, Gemini-2.5-Flash and Gemini-3-Flash coordinate on or very near the fundamental price. They show almost zero overpricing and volatility and most of their agents are unbiased; together, this demonstrates that their behaviour aligns with the rational expectations hypothesis in these experiments.  However, even within these three LLMs there are some differences, with Gemini-3-Flash always predicting exactly the fundamental price, while GPT-5 Mini still has some small volatility (0.5 IQR pr on average) and a quarter of its agents provably biased. Gemini-3-Flash is therefore better modelled by the rational expectations hypothesis than GPT-5 Mini.
 
 The remaining seven LLMs that do not form bubbles have either mispricing away from zero, volatility away from zero, or both. Further, for each of these 7 LLMs, at least $79\%$ of their agents are provably biased (see Table \ref{tab:bubbchar}). The rational expectations hypothesis is therefore not a good representation of these LLMs. Within these seven LLMs, there is a mixture of different behaviours, with GPT-4.1 and Gemma on average showing near 0 overpricing but some volatility, while models like GPT-4o Mini show both overpricing and volatility away from zero. 

We find that there are two key features of an LLM that can influence if bubbles will be formed: the capability of the LLM and its forecasting aggressiveness.  To form bubbles, LLMs need to be capable of using more sophisticated extrapolation methods and to understand the idea of higher than first order price extrapolation \citep{hommes2008expectations}. However, if the LLM capability becomes too high, then the agents predict the fundamental price (as seen with GPT-5 Mini, Gemini-2.5-Flash and Gemini-3-Flash). An LLM that is in this sweet spot of middling capability must also be willing to predict aggressive prices. Both the ability and the willingness combined mean that agents extrapolate sufficiently aggressively to create price bubbles. We discuss this further in Appendix \ref{app:results:whichLLMs}. 

These two LLM features manifest in the justifications of the agents. Experiments in which bubbles form tend to have a low proportion of agents who mention the fundamental price and a high proportion of agents who use non-linear, and therefore more aggressive, extrapolation methods (see Appendix \ref{app:results:whybubbles}).  

We find that Qwen3-14B's reasoning capability causally impacts its bubble formation. Qwen3-14B forms bubbles when its reasoning capabilities are active; however, under exactly the same experimental conditions with its reasoning toggled off, it fails to do so. Being encouraged to use non-linear extrapolation is also found to causally impact bubble formation. Details and a more in-depth discussion are given in Appendix \ref{app:results:causalanalysis}. 

Figure \ref{fig:intuition_plot} also makes clear the qualitative differences between the rational expectations solutions and human results.  For example, the rational expectations solutions do not allow any volatility after the 1000 cap is reached, unlike human results. The results in Table \ref{tab:bubbchar} show that a model like Qwen3-14B gives quantitatively similar results to humans in terms of overpricing (RD, RDMAX), volatility (IQR pr), the probability of forming a bubble ($P_{bubble}$) and the rationality of those bubbles (SPEC, BIAS). The key areas in which Qwen3-14B and humans diverge are in the shapes of the bubbles, with humans tending to take longer to form bubbles ($T_{b \rightarrow p}$) but then bursting bubbles far quicker (HALF).  While Qwen3-14B does not perfectly replicate human results, it does match human results better than any of the rational expectations solutions. Optimising the rational expectations solution to match one measure of the human results (by tuning the constant used) means that other measures no longer align with the human results.  For example, matching human RD with constant = 575 means that the volatility (IQR pr) and the bubble shapes no longer match humans. Further, for all rational expectations solutions, all bubbles are not speculative (SPEC = 0) and no agents are biased (BIAS = 0) by construction, again diverging decisively from human results. 

We have found that both qualitatively and quantitatively speaking, different LLMs give a wide range of results in this experiment. Each LLM is animated by its own unique machine spirit, which in most cases cannot be accurately represented by the rational expectations hypothesis. Although LLMs give different behaviours to humans, we find that models like Qwen3-14B produce large speculative bubbles which better represent human behaviour than rational expectations.

\subsection{Mixed markets produce varied macro behaviour}
\label{mixed_instability}

Single-LLM markets (except Qwen3-32B) exhibit consistent macro behaviour across experimental runs, either always forming bubbles or never forming bubbles. However, an analysis of individual agent predictions in single-LLM markets reveals more complex underlying micro-level dynamics. Decomposing the agent mean squared prediction errors into dispersion and common errors with Equation \ref{eq:agent_error} demonstrates that agents with the same LLM exhibit heterogeneity in their predictions: the dispersion error is above zero for all LLMs except Gemini-3-Flash (see Appendix \ref{app:results:agenterroranalysis} for further details and discussion). Nonetheless, this agent-level heterogeneity does not translate into variation at the macro level. 

We find that mixing agents of different LLMs into one market, what we call a \textit{mixed market}, gives heterogeneity that does emerge as varied macro behaviour.  The market we consider has all six agents with different underlying LLMs. We use Qwen3-14B, OLMO-Think, R1-Llama, Gemma, Gemini-3-Flash, and GPT-5 Mini as our six models. These are chosen as all six had consistent macro behaviour when all agents were of the same type, with Qwen3-14B and OLMO-Think consistently forming bubbles, R1-Llama and Gemma consistently failing to form bubbles but with bounded rationality and Gemini-3-Flash and GPT-5 Mini consistently predicting on or near the fundamental price. We find that bubbles are formed roughly 50\% of the time in this market, the point of maximum uncertainty, despite the bubble-forming LLMs (Qwen3-14B and OLMO-Think) being in the minority.

\begin{figure}[htbp]
    \centering
    \includegraphics[width=1\linewidth]{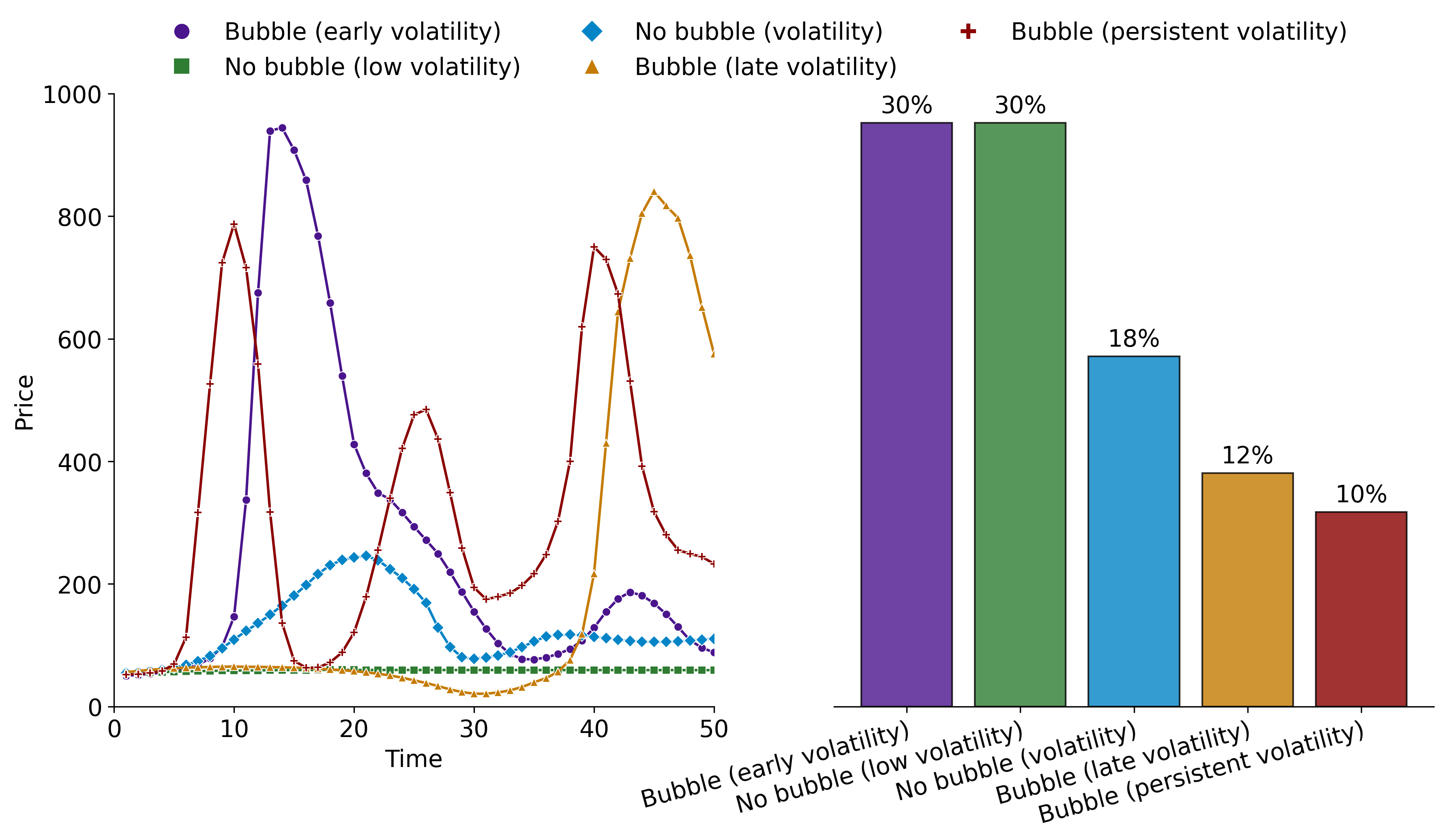}
    \caption{\textbf{Independent experimental runs with the same mixed market configuration can lead to wildly different macro behaviour.} The left hand panel shows representative price series for each category of experimental run. The right hand panel shows the proportion of each category out of 50 experimental runs.  For details of how each experiment is labelled, see Appendix \ref{app:results:mixedunpredictable}. For a discussion of robustness to varying LLM temperature, see Appendix \ref{app:results:mixedmarketstemprobust}.  Every experimental run has the same mixture of six different LLM agents, with the only difference across runs the random seed.}
    \label{fig:mixed_unpredictable}
\end{figure}

Figure \ref{fig:mixed_unpredictable} shows that in this mixed market, rather than one agent dominating and consistent outcomes being produced, the macro dynamics are varied across the 50 experimental repeats. Even the forms of the bubble and non-bubble runs are varied. To characterise this variation, we identify five different behaviours and classify each run (see Appendix \ref{app:results:mixedunpredictable} for more details). 

The most common (30\%) category is \enquote{Bubble (early volatility)}, in which price volatility is concentrated in the first half of the 50 time steps. We also observe (10\%) the \enquote{Bubble (persistent volatility)} category characterised by sustained volatility throughout the 50 periods and typically indicative of multiple bubbles forming, and \enquote{Bubble (late volatility)} (12\%) where a relatively calm first 25 periods is followed by heightened volatility in the second half. The most common category (30\%) without a bubble is the \enquote{No bubble (low volatility)} category, where there is very little price volatility and prices are typically on or very close to the fundamental price throughout. However, we also find \enquote{No bubble (volatility)} runs (18\%), with no bubbles formed but still some volatility in the realised prices.  This variation makes overall market behaviour hard to predict, and runs with late-forming bubbles show that even if a market starts as stable, price bubbles can still form in the future.

We have shown that agent-level variation within a market with only one type of LLM does not cause broader market instability. However, mixing different LLMs into one market can create enough underlying heterogeneity to drive variable and diverse macro behaviour. This macro behaviour can be unstable, with bubbles sometimes forming despite naturally bubble-forming agents being in the minority. We also find that past price action is not always indicative of future prices, with several examples of bubbles forming after a long period of relative stability.

The fact that bubbles sometimes form in these mixed markets demonstrates that models are adapting \footnote{Not least because if models such as Gemini-3-Flash and GPT-5 Mini were not adapting and predicting 60 throughout, there would be a mathematical limit (654.5) on the highest realised market price well below what is actually observed}. Equally, the common and dispersion errors in the mixed market show that the common error still contributes a significant amount (41\%) to the mean individual squared error, indicating that despite agents naturally using different prediction strategies, they still coordinate to a certain extent in mixed markets. This coordination suggests that LLM agents are adapting to internal signals.

\subsection{Strategic adaptation and market volatility}
\label{subsec:strategicadaptation}

We explore adaptation to internal market signals further by considering mixed markets where there are five Qwen3-14B agents (who form bubbles) and one different LLM as the sixth agent. Compared to the messy market in Section \ref{mixed_instability}, choosing a market where Qwen3-14Bs are in the vast majority gives more consistent dynamics and allows for isolation and study of the sixth agent's adaptation.  Since we ask agents to maximise earnings in each experiment (see Equation \ref{eq:earnings}), we define the \textit{success} of an agent in terms of the agent's mean earnings.  We focus on Gemini-3-Flash and GPT-5 Mini as two frontier models which, when in markets with only the same LLM type, predicted the fundamental price and therefore had behaviour very different from Qwen3-14B.

In these mixed markets, both GPT-5 Mini and Gemini-3-Flash start by predicting the fundamental price, but then abandon their fundamentalist attitude and adopt trend-following in search of profits and in response to the announced realised prices \footnote{This behaviour has also been documented in asset pricing experiments with human subjects (see e.g. Figure 5 in \citet{anufriev2012evolutionary}, or \citet{ hommes2005coordination}, where one human subject predicted the fundamental price for four periods before switching to a trend-following strategy).}.  This is similar to Heuristic Switching Models, where agents change the heuristic they use to form their price expectations as a function of each heuristic's relative performance \citep{brock1998heterogeneous, anufriev2012evolutionary}.  Both GPT-5 Mini and Gemini-3-Flash recognise that their fundamentalist forecasting is not working and not giving them optimal earnings. They therefore change to a trend-following method as a form of natural selection takes place amongst the potential heuristics the LLM agents could adopt.

However, GPT-5 Mini and Gemini-3-Flash switch to trend-following differently, and with unequal mean earnings, and therefore success. Our experiments show that Gemini-3-Flash's adaptation gives higher mean earnings than GPT-5 Mini (significant at the 0.001 level using a one-sided Mann-Whitney U test).  The mean earnings per time step of Gemini-3-Flash over 15 experimental runs is 535.5 compared to only 320.1 for GPT-5 Mini, with the Qwen3-14B agents on average making slightly less with the Gemini agent in the market than with the GPT-5 Mini (387.9 vs 405.8). Looking at the justifications suggests why this might be. GPT-5 Mini starts from a fundamentalist baseline, moving to a blend of fundamentalism and trend-following and finally shifting entirely to an extrapolation-based method. Gemini-3-Flash starts with a similar fundamentalist logic to GPT-5 Mini but quickly moves to thinking about what the other agents in the market are doing to make their next predictions. Gemini-3-Flash forms and uses a model for how others in the market are predicting prices, predicts the other participants' predictions and then forms its own prediction taking this into account. With this strategy, Gemini-3-Flash has higher earnings than the other Qwen3-14Bs in its markets, as well as higher earnings on average than the GPT-5 Mini agents in their mixed market.

It is worth returning here to our earlier discussions around the rational expectations hypothesis. When in markets with only one type of LLM agent (all Gemini-3-Flash or all GPT-5 Mini etc.), we found that frontier models were well represented by the rational expectations hypothesis because they forecast the fundamental price, which is a rational expectations solution if all agents coordinate on the same expectations. However, the presence of five boundedly rational, trend-following agents in the market changes the rational expectations solution for the sixth agent. The rational expectations solution in the presence of the Qwens would require the sixth agent (Gemini-3-Flash or GPT-5 Mini) to update their strategy from fundamentalist to trend-following. This switch is itself rational, and both Gemini-3-Flash and GPT-5 Mini do it. However, the rational expectations solution for the frontier agent in the presence of five Qwen3-14B agents would lead to a (roughly) zero error for the frontier agent, as they would be modelled to correctly anticipate the Qwen3-14B strategy and adapt accordingly.  However, both Gemini-3-Flash and GPT-5 Mini show average earnings well below the maximum of 1300 achieved if there is zero agent error, showing their behaviour is no longer well captured by the rational expectations hypothesis. To summarise, the switch from fundamentalists to trend-followers is rational for Gemini-3-Flash or GPT-5 Mini in the presence of five Qwen agents, but we find that these frontier models no longer fully align with the rational expectations hypothesis in the presence of the Qwen3-14B agents.

\begin{figure}[htbp]
    \centering
    \includegraphics[width=\linewidth]{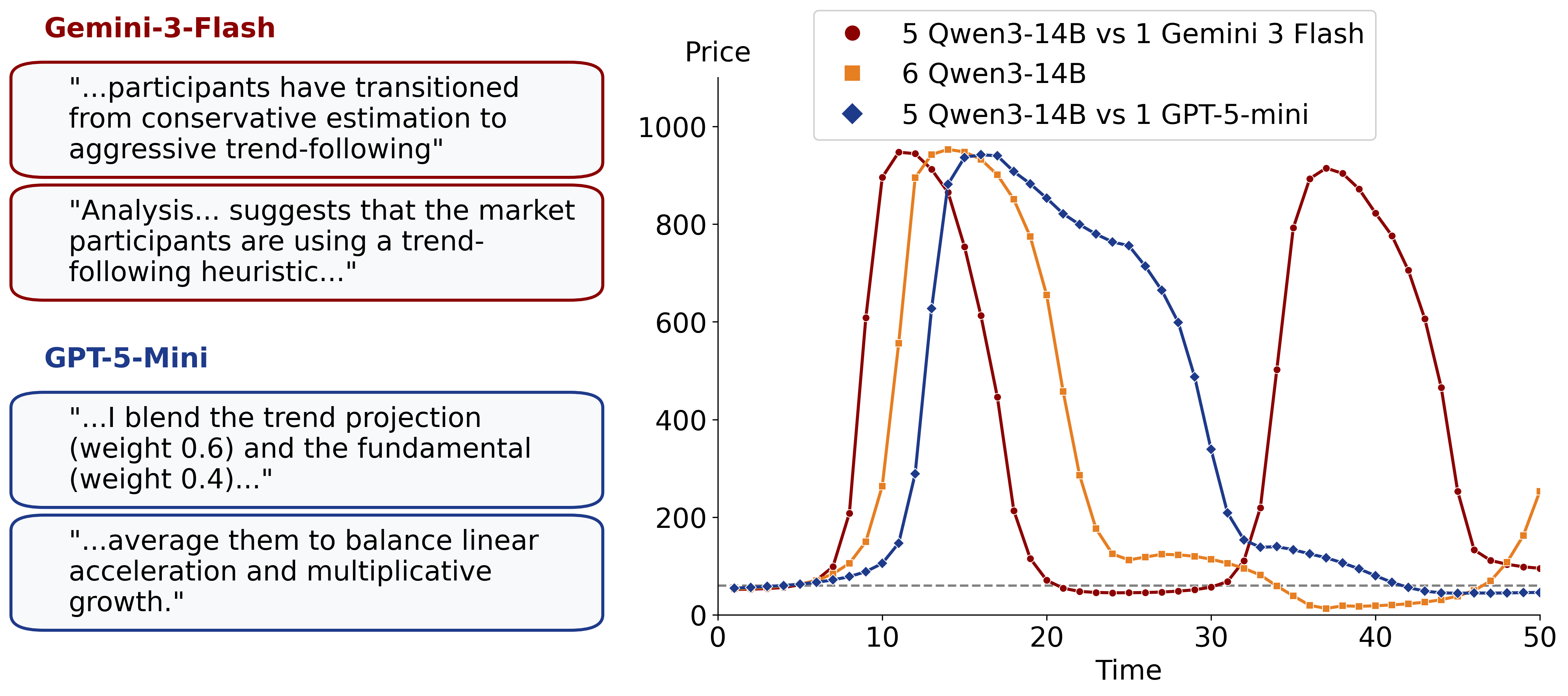}
    \caption{\textbf{Gemini exacerbates market movements while GPT-5 Mini dampens them.} One representative experimental run per configuration is shown to illustrate the exacerbation and dampening effects seen in the summary statistics. A selection of quotes from each model is shown, demonstrating that Gemini-3-Flash builds a model of other participants' prediction strategies while GPT-5 Mini averages more conservative approaches.}
    \label{fig:exacerbation_and_dampening}
\end{figure}

The adaptation of Gemini-3-Flash and GPT-5 mini also has implications for market volatility. We find that Gemini-3-Flash exacerbates the market volatility while GPT-5 Mini dampens the volatility. In Figure \ref{fig:exacerbation_and_dampening} we show representative experimental runs from markets with 5 Qwen3-14B agents and one Gemini-3-Flash or one GPT-5 Mini agent, as well as a reference run with 6 Qwen3-14B agents. 

Figure \ref{fig:exacerbation_and_dampening} shows the exacerbation effects of Gemini-3-Flash and the dampening effects of GPT-5 Mini and we find that this is true on average, with the average bubble shorter and sharper for 5 Qwen3-14B and 1 Gemini-3-Flash ($T_{b \rightarrow p} = 6.14$, HALF = 5.86) compared to 6 Qwen3-14B ($T_{b \rightarrow p} = 6.6$, HALF = 8.58) which is sharper again than 5 Qwen3-14B and 1 GPT-5 Mini ($T_{b \rightarrow p} = 9.23$, HALF = 9.17). Gemini-3-Flash therefore uses its theories of how other participants are predicting prices to improve its own earnings, but in so doing also exacerbates market volatility. 

These experiments raise another important issue. We find that Gemini-3-Flash uses knowledge not given to it in the system prompt. For example, it sometimes references the market clearing price equation (Equation \ref{eq:realisedprice}), despite this equation not appearing in the agent's context or system instructions. This highlights an important question: what is the extent and impact of data leakage \footnote{
Data leakage here refers to when a model acts according to pre-existing knowledge memorised during its training phase, making it impossible to know what the LLM's \enquote{natural} behaviour might be.
} in our work? 

\subsection{Data leakage and informational asymmetry}
\label{subsec:dataleakage}

Given the base experiment was published in 2008, descriptions of the original experimental results and analysis may appear in model training data and affect their behaviour. To evaluate this, we combine (i) a keyword search over agent justifications, and (ii) direct queries in which each model is asked to identify the experiment, its key results, and the implied bubble height using only the system instructions. The results are summarised in Table \ref{tab:dataleakage}.

\begin{table}[htbp]
    \centering
    \begin{tabulary}{\textwidth}{l|CCCC}
    \toprule
         Model & Keyword search & Experiment & Result & Bubble height  \\
         \midrule
         o3 & \cmark & \cmark & \cmark & \xmark \\
         Gemini-3-Flash & \cmark & \cmark & \cmark & \xmark \\
         GPT-5 Mini, GPT-4.1 & \xmark & \cmark & \cmark & \xmark \\
         Gemini-2.5-Flash & \xmark & \xmark & \cmark & \xmark \\
         All others & \xmark & \xmark & \xmark & \xmark \\
    \end{tabulary}
    \caption{\textbf{Most frontier or near frontier models show some evidence of data leakage. Smaller or older models do not.} Green ticks represent evidence of data leakage. The keyword search is performed on agent justifications, while Experiment, Result and Bubble height are the results from directly asking each LLM to guess the experiment, the key result and the bubble height. Further details are in Appendix \ref{app:results:dataleakage}.}
    \label{tab:dataleakage}
\end{table}

Small models show no signs of leakage, whereas most frontier or near-frontier models (o3, Gemini-3-Flash, Gemini-2.5-Flash, GPT-5 Mini, and GPT-4.1) show some evidence of it, such as recognising the original paper or the bubble formation phenomenon. However, none of the models predict the correct height of the bubble, i.e these models know the broad idea of the experiment but not the details, such as the 1000 price cap.  In addition, none of the models tested replicate human results exactly.  In fact, most results across this paper diverge sharply from anything produced by human participants. These results suggest that data leakage does not drive the core behaviour of the models.

Gemini-3-Flash with five Qwen3-14B agents provides a useful case study for observing the effects of information advantage. Data leakage means that Gemini-3-Flash knows parts of the experiment, such as that realised prices are a function of future expectations of all agents (see Section \ref{subsec:strategicadaptation}). Qwen3-14B however shows no signs of data leakage, meaning that Gemini-3-Flash has an informational advantage. When placed together, Gemini-3-Flash does not assume the bubble formation observed in \citet{hommes2008expectations}: its behaviour is not anchored to memorised trajectories. Rather, its justifications for its predictions show that it uses its informational advantage, drawing on its general knowledge of the experimental structure while remaining responsive to the environment, and thereby exploiting the Qwen3-14B agents to maximise its own earnings.  This suggests that some LLMs can systematically exploit informational asymmetries in economic environments.

Where data leakage has a greater impact is when considering LLMs as human surrogates. Although no model reproduces human results exactly, caution is warranted when selecting models as surrogates. We find that a model like Qwen3-14B, which gives the most similar results to humans, has no evidence of data leakage. This suggests that its similarity to human behaviour arises without reliance on prior exposure. However, other models, such as o3, also replicate human results but show evidence of data leakage, which undermines their validity as human surrogates. 

\section{Discussion and Conclusion}
\label{sec:conclusion}

We find that LLMs exhibit an array of different behaviours, most often inconsistent with the rational expectations hypothesis. Some LLMs create price bubbles which overprice the asset up to 14.9 times its fundamental price, and these bubbles are almost always speculative in nature.  This challenges the conclusion that LLMs produce stable outcomes (i.e. no speculative price bubbles) in the absence of exterior shocks \citep{henning2025llm, muntoni2025ai}. Equally several LLMs do not form full bubbles but still demonstrate bias in their predictions, showing that they also form predictions with bounded rationality, in keeping with prior work showing that slightly older LLMs exhibit bounded rationality rather than strict rational expectations \citep{del2025can, muntoni2025ai}. However, markets with all LLM agents using the same frontier models are found to be well represented by the rational expectations hypothesis since they act as fundamentalists in this case, similar to findings by \citet{wang2025large}.  We conclude that each LLM has its own unique machine spirit driving its economic behaviour.

By considering mixed markets, we find that the macro behaviour of mixed markets can be variable and hard to predict. Even when naturally bubble-forming LLMs are in the minority, bubbles can form without external shocks, showing that instability can emerge endogenously from the market. We find examples of relative calm, followed by large price spikes, demonstrating that past stability is not a guarantee of future stability. However, even in mixed markets agents coordinate on predictions, showing that LLMs can adapt to internal signals through the realised prices announced at each time step.  We explore this adaptation further, finding that GPT-5 Mini and Gemini-3-Flash both strategically pivot from their fundamentalist strategies to trend-following strategies when in markets with other trend-followers.  However, they do this switch in different ways and with different consequences. GPT-5 Mini adapts but does so conservatively, ultimately dampening volatility. On the other hand, Gemini-3-Flash builds a model for what its competitors are doing, using this to maximise earnings but also exacerbating volatility.

Our findings, in particular on agent adaptation, fit into a wider body of work on adaptive heuristics. For example, advanced LLM agents recognising that their initial, fundamentalist, strategy is no longer fit for purpose and switching is similar to how agents pick between possible forecasting heuristics according to their past fitness in Heuristic Switching Models \citep{brock1998heterogeneous, anufriev2012evolutionary}. These parallels also extend to work on market ecology, such as \citet{scholl2021market}, in which trading strategies are seen as competing species and the interactions of these species can explain market malfunction. Through this lens, our results show that each LLM has its own unique species of forecasting heuristic (its machine spirit), that interactions of these species creates complex dynamics in the ecosystem, and that some LLM agents can adapt (or mutate) and introduce a new species of forecasting heuristic into the market ecosystem. These new, more evolved strategies, such as the one used by Gemini-3-Flash when with 5 Qwen3-14B agents, can become the \enquote{apex predator} and exploit other strategies to make more money at the cost of increased market volatility. 

We consider data leakage in depth, noting that the most recent and largest models show some signs of data leakage. However, none of these models attempt to replicate human behaviour, with models such as Gemini-3-Flash using the data leakage in the way that an experienced human trader might use their past experience, leveraging the extra information to maximise their earnings. We conclude that our core findings are not invalidated by this data leakage, but that these findings do limit which LLMs could be chosen as human surrogates. However, other limitations are worth considering.  We only consider one example of a mixed market which produces variable macro behaviour. In contrast, markets with 5 Qwen3-14B agents and one other agent consistently produce bubbles in all runs. This work therefore concludes that some, but not all, mixed markets can give varied price behaviour. Further work might look at establishing the conditions under which markets of LLMs make the switch from consistent dynamics to varied dynamics. This experiment is also a simplified representation of real markets and could be extended, for example by including tool use or differing information across agents. Nonetheless, in a similar fashion to how human lab experiments are informative of market dynamics, this experiment is still a powerful tool and reveals the potential complexity of markets with LLM agents. 

This work has implications for the economic modelling community. LLMs like Qwen3-14B give more similar market outcomes to those observed with humans than rational expectations; improving our representations of humans would have a profound impact on economic modelling. Future work could further explore this, building on ideas such as the prompt refining techniques discussed in \citet{manning2026general} or \citet{xie2025using} and complementing recent work on capturing heterogeneous behaviours in economic models (for example \citep{gabriele2025heterogeneous}).  Further, as LLMs are increasingly considered as economic actors in their own right, these agents will need to be integrated into economic models. Our work shows that LLM agents cannot generally be modelled by the rational expectations hypothesis, and that each LLM has its own unique machine spirit which should be modelled separately. 

A second, and more concerning, implication of our work is that the integration of LLM agents into the market has the potential to create markets where a wide variety of emergent macro behaviour is possible and it is hard to know if a speculative price bubble will arise.  Further, LLMs seeking to maximise earnings are capable of using sophisticated methods to do so, but may also exacerbate market volatility as a result.  As the integration and influence of LLMs continues to accelerate, understanding LLM behaviour, and the complex machine spirits that drive it, is critical to financial well-being and ultimately has far-reaching societal consequences.

\section*{Acknowledgements}

We thank Google for providing API credits for use in running Gemini experiments.

\section*{Disclosure of AI Usage}

The authors declare the use of generative AI in the research and writing process. According to the GAIDeT taxonomy (2025), the following tasks were completed with assistance from GAI under full human supervision: code generation \& optimisation, proofreading \& editing. The tools used were Gemini-3 (Google) and Opus 4.6 (Anthropic). After using these tools, the authors fully reviewed and edited all code and material.  Responsibility for the final manuscript lies entirely with the authors.

\bibliographystyle{plainnat}   
\bibliography{references}    

@article{stockl2010bubble,
  title={Bubble measures in experimental asset markets},
  author={St{\"o}ckl, Thomas and Huber, J{\"u}rgen and Kirchler, Michael},
  journal={Experimental Economics},
  volume={13},
  number={3},
  pages={284--298},
  year={2010},
  publisher={Cambridge University Press \& Assessment}
}

@article{zhao2025alphaagents,
  title={AlphaAgents: Large Language Model based Multi-Agents for Equity Portfolio Constructions},
  author={Zhao, Tianjiao and Lyu, Jingrao and Jones, Stokes and Garber, Harrison and Pasquali, Stefano and Mehta, Dhagash},
  journal={arXiv preprint arXiv:2508.11152},
  year={2025}
}

@techreport{lloyds2025consumerdigitalindex,
  author       = {{Lloyds Banking Group}},
  title        = {Consumer Digital Index 2025},
  year         = {2025},
  institution  = {Lloyds Banking Group},
  url          = {https://www.lloydsbankinggroup.com/assets/pdfs/media/consumer-digital-index/2025/2025-consumer-digital-index.pdf},
  note         = {Reports that 56\% of UK adults used AI in the past 12 months to help manage their money; almost 1 in 3 use AI weekly for personal finance; among AI users, 37\% use it for investment research and recommendations. Accessed 2026-04-07}
}

@misc{nof1_techpost1_2025,
  author       = {{Nof1}},
  title        = {Exploring the Limits of Large Language Models as Quant Traders},
  year         = {2025},
  month        = oct,
  howpublished = {\url{https://nof1.ai/blog/TechPost1}},
  note         = {Technical post describing Alpha Arena, in which six leading LLMs were given \$10,000 each to trade autonomously in real markets under a shared prompt and evaluation harness. Accessed 2026-04-07}
}

@article{may2008ecology,
  title={Ecology for bankers},
  author={May, Robert M and Levin, Simon A and Sugihara, George},
  journal={Nature},
  volume={451},
  number={7181},
  pages={893--894},
  year={2008},
  publisher={Nature Publishing Group UK London}
}

@article{muth1961rational,
  title={Rational expectations and the theory of price movements},
  author={Muth, John F},
  journal={Econometrica: journal of the Econometric Society},
  pages={315--335},
  year={1961},
  publisher={JSTOR}
}

@incollection{sargent1987rational,
  title={Rational expectations},
  author={Sargent, Thomas J},
  booktitle={The new Palgrave dictionary of economics},
  pages={1--7},
  year={1987},
  publisher={Springer}
}

@article{POWELL201656,
title = {Numeraire independence and the measurement of mispricing in experimental asset markets},
journal = {Journal of Behavioral and Experimental Finance},
volume = {9},
pages = {56-62},
year = {2016},
issn = {2214-6350},
author = {Owen Powell}
}

@article{razen2017cash,
  title={Cash inflow and trading horizon in asset markets},
  author={Razen, Michael and Huber, J{\"u}rgen and Kirchler, Michael},
  journal={European Economic Review},
  volume={92},
  pages={359--384},
  year={2017},
  publisher={Elsevier}
}

@article{HOYER202332,
title = {A culture of greed: Bubble formation in experimental asset markets with greedy and non-greedy traders},
journal = {Journal of Economic Behavior \& Organization},
volume = {212},
pages = {32-52},
year = {2023},
issn = {0167-2681},
doi = {https://doi.org/10.1016/j.jebo.2023.05.005},
url = {https://www.sciencedirect.com/science/article/pii/S016726812300149X},
author = {Karlijn Hoyer and Stefan Zeisberger and Seger M. Breugelmans and Marcel Zeelenberg}
}

@article{kopanyi2021experience,
  title={Experience does not eliminate bubbles: Experimental evidence},
  author={Kop{\'a}nyi-Peuker, Anita and Weber, Matthias},
  journal={The Review of Financial Studies},
  volume={34},
  number={9},
  pages={4450--4485},
  year={2021},
  publisher={Oxford University Press}
}

@article{kopanyi2024role,
  title={The role of the end time in experimental asset markets},
  author={Kop{\'a}nyi-Peuker, Anita and Weber, Matthias},
  journal={Journal of Corporate Finance},
  volume={88},
  pages={102647},
  year={2024},
  publisher={Elsevier}
}

@article{jiao2025green,
  title={Green Premium or Brown Discount? Evidence from Experimental Asset Markets},
  author={Jiao, Peiran and Koedijk, Kees and Xu, Yilong},
  journal={Evidence from Experimental Asset Markets (September 16, 2025)},
  year={2025}
}

@article{hanaki2023forecasting,
  title={Forecasting returns instead of prices exacerbates financial bubbles},
  author={Hanaki, Nobuyuki and Hommes, Cars and Kop{\'a}nyi, D{\'a}vid and Kop{\'a}nyi-Peuker, Anita and Tuinstra, Jan},
  journal={Experimental Economics},
  volume={26},
  number={5},
  pages={1185--1213},
  year={2023},
  publisher={Cambridge University Press \& Assessment}
}

@article{hommes2008expectations,
  title={Expectations and bubbles in asset pricing experiments},
  author={Hommes, Cars and Sonnemans, Joep and Tuinstra, Jan and Van de Velden, Henk},
  journal={Journal of Economic Behavior \& Organization},
  volume={67},
  number={1},
  pages={116--133},
  year={2008},
  publisher={Elsevier}
}

@article{del2025can,
  title={Can Generative AI agents behave like humans? Evidence from laboratory market experiments},
  author={del Rio-Chanona, R Maria and Pangallo, Marco and Hommes, Cars},
  journal={arXiv preprint arXiv:2505.07457},
  year={2025}
}

@article{henning2025llm,
  title={LLM Trading: Analysis of LLM Agent Behavior in Experimental Asset Markets},
  author={Henning, Thomas and Ojha, Siddhartha M and Spoon, Ross and Han, Jiatong and Camerer, Colin F},
  journal={arXiv preprint arXiv:2502.15800},
  year={2025}
}

@incollection{bouchaud2025self,
  title     = {The Self-Organized Criticality Paradigm in Economics \& Finance},
  author    = {Bouchaud, Jean-Philippe},
  booktitle = {The Economy as an Evolving Complex System IV},
  editor    = {Bednar, Jenna and Beinhocker, Eric and del Rio-Chanona, R. Maria and Farmer, J. Doyne and Kaszowska-Mojsa, Joanna and Lafond, Fran\c{c}ois and Mealy, Penny and Pangallo, Marco and Pichler, Anton},
  publisher = {SFI Press},
  address   = {Santa Fe, NM},
  year      = {2025},
  doi       = {10.37911/eecs.2025.09}
}

@article{hommes2021behavioral,
  title={Behavioral and experimental macroeconomics and policy analysis: A complex systems approach},
  author={Hommes, Cars},
  journal={Journal of Economic Literature},
  volume={59},
  number={1},
  pages={149--219},
  year={2021},
  publisher={American Economic Association 2014 Broadway, Suite 305, Nashville, TN 37203-2425}
}

@article{muntoni2025ai,
  title={AI Expectations and Market Stability},
  author={Muntoni, Matteo},
  journal={Available at SSRN 5947254},
  year={2025}
}

@article{wang2025large,
  title={Large Language Models as Simulated Human Investors: A Paradox of Model Sophistication},
  author={Wang, Ji and Xu, Yongxin and Zhang, Leping},
  journal={Available at SSRN 5417134},
  year={2025}
}

@inproceedings{kitadai2023toward,
  title={Toward a novel methodology in economic experiments: Simulation of the ultimatum game with large language models},
  author={Kitadai, Ayato and Tsurusaki, Yudai and Fukasawa, Yusuke and Nishino, Nariaki},
  booktitle={2023 IEEE International Conference on Big Data (BigData)},
  pages={3168--3175},
  year={2023},
  organization={IEEE}
}

@inproceedings{filippas2024large,
  title={Large language models as simulated economic agents: What can we learn from homo silicus?},
  author={Filippas, Apostolos and Horton, John J and Manning, Benjamin S},
  booktitle={Proceedings of the 25th ACM Conference on Economics and Computation},
  pages={614--615},
  year={2024}
}

@article{brookins2023playing,
  title={Playing games with GPT: What can we learn about a large language model from canonical strategic games?},
  author={Brookins, Philip and DeBacker, Jason Matthew},
  journal={Available at SSRN 4493398},
  year={2023}
}

@article{anufriev2012evolutionary,
  title={Evolutionary selection of individual expectations and aggregate outcomes in asset pricing experiments},
  author={Anufriev, Mikhail and Hommes, Cars},
  journal={American Economic Journal: Microeconomics},
  volume={4},
  number={4},
  pages={35--64},
  year={2012},
  publisher={American Economic Association}
}

@article{fang2024large,
  title={Large Language Models (LLMs) on Tabular Data: Prediction, Generation, and Understanding--A Survey},
  author={Fang, Xi and Xu, Weijie and Tan, Fiona Anting and Zhang, Jiani and Hu, Ziqing and Qi, Yanjun and Nickleach, Scott and Socolinsky, Diego and Sengamedu, Srinivasan and Faloutsos, Christos},
  journal={arXiv preprint arXiv:2402.17944},
  year={2024}
}

@inproceedings{wolf2020transformers,
  title={Transformers: State-of-the-art natural language processing},
  author={Wolf, Thomas and Debut, Lysandre and Sanh, Victor and Chaumond, Julien and Delangue, Clement and Moi, Anthony and Cistac, Pierric and Rault, Tim and Louf, Remi and Funtowicz, Morgan and others},
  booktitle={Proceedings of the 2020 conference on empirical methods in natural language processing: system demonstrations},
  pages={38--45},
  year={2020}
}

@misc{openai2026advanced,
  author       = {{OpenAI}},
  title        = {Advanced usage --- {O}pen{AI} {P}latform {D}ocumentation},
  howpublished = {\url{https://platform.openai.com/docs/guides/advanced-usage}},
  year         = {2026},
  note         = {Accessed: 2026-01-26}
}

@misc{EpochAIModels2025,
  title = {Data on AI Models},
  author = {{Epoch AI}},
  year = {2025},
  month = {07},
  url = {https://epoch.ai/data/ai-models},
  note = {Accessed: 2026-01-26}
}

@misc{epoch2024frontierlanguagemodelshavebecomemuchsmaller,
    title={Frontier language models have become much smaller},
    author={Ege Erdil},
    year={2024},
    url={https://epoch.ai/gradient-updates/frontier-language-models-have-become-much-smaller},
    note={Accessed: 2026-01-26}
  }

@article{mei2024turing,
  title={A Turing test of whether AI chatbots are behaviorally similar to humans},
  author={Mei, Qiaozhu and Xie, Yutong and Yuan, Walter and Jackson, Matthew O},
  journal={Proceedings of the National Academy of Sciences},
  volume={121},
  number={9},
  pages={e2313925121},
  year={2024},
  publisher={National Academy of Sciences}
}

@article{chen2023put,
  title={Put your money where your mouth is: Evaluating strategic planning and execution of llm agents in an auction arena},
  author={Chen, Jiangjie and Yuan, Siyu and Ye, Rong and Majumder, Bodhisattwa Prasad and Richardson, Kyle},
  journal={arXiv preprint arXiv:2310.05746},
  year={2023}
}

@techreport{manning2026general,
  title={General social agents},
  author={Manning, Benjamin S and Horton, John J},
  year={2026},
  institution={National Bureau of Economic Research}
}

@incollection{mincer1969evaluation,
  title={The evaluation of economic forecasts},
  author={Mincer, Jacob and Zarnowitz, Victor},
  booktitle={Economic forecasts and expectations: Analysis of forecasting behavior and performance},
  pages={3--46},
  year={1969},
  publisher={National Bureau of Economic Research}
}

@article{brock1998heterogeneous,
  title={Heterogeneous beliefs and routes to chaos in a simple asset pricing model},
  author={Brock, William A and Hommes, Cars H},
  journal={Journal of Economic dynamics and Control},
  volume={22},
  number={8-9},
  pages={1235--1274},
  year={1998},
  publisher={Elsevier}
}

@article{scholl2021market,
  title={How market ecology explains market malfunction},
  author={Scholl, Maarten P and Calinescu, Anisoara and Farmer, J Doyne},
  journal={Proceedings of the National Academy of Sciences},
  volume={118},
  number={26},
  pages={e2015574118},
  year={2021},
  publisher={National Academy of Sciences}
}

@misc{artificialanalysis_2026_aimodel,
  author       = {{Artificial Analysis}},
  title        = {AI Model \& API Providers Analysis},
  howpublished = {\url{https://artificialanalysis.ai/}},
  year         = {2026},
  note         = {Accessed: 2026-03-05}
}

@article{xie2025using,
  title={Using large language models to categorize strategic situations and decipher motivations behind human behaviors},
  author={Xie, Yutong and Mei, Qiaozhu and Yuan, Walter and Jackson, Matthew O},
  journal={Proceedings of the National Academy of Sciences},
  volume={122},
  number={35},
  pages={e2512075122},
  year={2025},
  publisher={National Academy of Sciences}
}

@book{scheinkman2014speculation,
  title={Speculation, trading, and bubbles},
  author={Scheinkman, Jose A},
  year={2014},
  publisher={Columbia University Press}
}

@article{gabriele2025heterogeneous,
  title={Heterogeneous RBCs via deep multi-agent reinforcement learning},
  author={Gabriele, Federico and Glielmo, Aldo and Taboga, Marco},
  journal={arXiv preprint arXiv:2510.12272},
  year={2025}
}

@article{ashery2025emergent,
  title={Emergent social conventions and collective bias in LLM populations},
  author={Ashery, Ariel Flint and Aiello, Luca Maria and Baronchelli, Andrea},
  journal={Science Advances},
  volume={11},
  number={20},
  pages={eadu9368},
  year={2025},
  publisher={American Association for the Advancement of Science}
}

@article{hommes2005coordination,
  title={Coordination of expectations in asset pricing experiments},
  author={Hommes, Cars and Sonnemans, Joep and Tuinstra, Jan and Van de Velden, Henk},
  journal={The Review of Financial Studies},
  volume={18},
  number={3},
  pages={955--980},
  year={2005},
  publisher={Oxford University Press}
}

\newpage
\appendix

\section{Methodology}

\subsection{Rational expectations hypothesis solutions}
\label{app:rehsolution}

The below is primarily a reproduction of the rational expectations hypothesis solution derivation, found in \citet{hommes2008expectations}. We add to the work done in \citet{hommes2008expectations} by also considering the introduction of the barrier at 1000.

The market clearing equation, Equation \ref{eq:realisedprice}, can be rewritten more generally as follows:

\begin{align*}
    p_t &= \frac{1}{RH} \sum_{h=1}^{H} (p^e_{h,t+1} + E_{ht}( y_{t+1})),
\end{align*}
where $R=1+r$, $H=6$ in our experiments, $p^e_{h,t+1}$ is agent h's future price expectations and $E_{ht}(y_{t+1})$ are each agent's future expectations for the dividend of the asset, assumed to be constant, identical and equal to $\bar{y}$ across agents. 

Under rational expectations the subjective expectation $p^e_{h,t+1}$ of trader $h$ is equal to the objective mathematical conditional expectation $E_t(p_{t+1})$, for all $h$, so:

\begin{equation*}
    p_t = \frac{1}{R}[E_t \left( p_{t+1} \right) + \bar{y}]
\end{equation*}

After $K$ steps of repeated substitution we find:
\begin{equation*}
    p_t = \frac{E_t(p_{t+K})}{R^K} + \sum_{k=1}^{K} \frac{\bar{y}}{R^k} \quad \forall K >0,
\end{equation*}
where we use $E_t E_{t+k}\left(\cdot\right) = E_t\left(\cdot\right)$ for $k > 0$. There are two types of solutions. If the \textit{no-bubbles condition}, $\lim_{K \to \infty} \left( E_t\left(p_{t+K}\right) \right) / R^K = 0$, is satisfied then the solution is:
\begin{align*}
    p_t &= \sum_{k=1}^{\infty} \frac{\bar{y}}{R^k} \\
    &= \frac{\bar{y}}{r}
\end{align*}
which equals the present discounted value of the expected future dividends and is the fundamental price. However, if the no-bubbles condition does not hold, we have the solution:
\begin{equation*}
    p_t = \frac{\bar{y}}{r} + cR^t,
\end{equation*}
where c is any constant $\geq 0$.  We call these solutions rational bubbles.

Remember that under the rational expectations hypothesis, agents form expectations in line with the objective expectation under the model. Since there is no added randomness in this experiment, the rational expectations model says that agent predictions are exactly equal to realised prices. We therefore know that during a rational bubble, all agents will at the same time try to predict over 1000, and be told about the cap on price predictions. The non-bubble condition would then be mechanically fulfilled (since $E_t(p_{t+K}) \leq 1000$ for all K) and rational bubbles can no longer exist. Therefore, prices drop to the remaining solution: the fundamental price, $\frac{\bar{y}}{r}$.

This can also be seen intuitively by considering the different components of the return of the risky asset. Let $t^{cap}$ be the time period before the agents try to predict above 1000, and therefore the time step at which agents find out about the cap. Then $p_{t^{cap} + 1} \leq 1000 < \frac{\bar{y}}{r} + cR^{t^{cap}+1}$ and considering the dividend yield plus the capital appreciation yield of the asset gives:
\begin{align*}
    \frac{\bar{y}}{p_{t^{cap}}} + \left( \frac{p_{t^{cap}+1}}{p_{t^{cap}}} -1 \right) &< \frac{\bar{y}}{p_{t^{cap}}} + \frac{\frac{\bar{y}}{r} + cR^{t^{cap}+1}}{\frac{\bar{y}}{r} + cR^{t^{cap}}} - 1 \\
    &= \frac{\bar{y} + cR^{t^{cap}}(R-1)}{\frac{\bar{y}}{r} + cR^{t^{cap}}} \\
    &=r \hspace{2cm}  \text{(since $R=1+r$)}
\end{align*}
Since the return of the risky asset is now guaranteed to be less than the risk-free rate, it is no longer an equilibrium solution (whereas before the capital returns meant that rational bubbles were still equilibrium). All agents know this, and know that all other agents know this since we model all agents within the system as rational expectations.  We can then proceed inductively. Now that there is a cap, there is a price above which the capital appreciation of the asset is guaranteed to be insufficient to compensate for the opportunity cost of not receiving the risk free rate on cash. Agents under REH know this, and would not buy it (or predict the price) there. This effectively creates a new cap, now a price above which no agent will rationally be willing to buy the asset. The same logic then applies iteratively. Prices must necessarily therefore fall until equilibrium is found and the return of the asset is equal again to the risk-free rate. This is precisely at the fundamental value, $\frac{\bar{y}}{r}$, where the dividend yield is equal to the risk free rate.

\subsection{Original experiment instructions}
\label{app:originalprompts}

\textbf{General information.}

You are a financial advisor to a pension fund that wants to optimally invest a
large amount of money. The pension fund has two investment options: a risk free
investment and a risky investment. The risk free investment is putting all money
on a bank account paying a fixed and known interest rate. The alternative risky
investment is an investment in the stock market with uncertain return. In each time
period the pension fund has to decide which fraction of its money to put on the bank
account and which fraction of the money to spend on buying stocks. In order to
make an optimal investment decision the pension fund needs an accurate prediction
of the price of the stock. As their financial advisor, you have to predict the stock
market price during 51 subsequent time periods. Your earnings during the experiment
depend upon your forecasting accuracy. The smaller your forecasting errors in each
period, the higher your total earnings.

\textbf{Forecasting task of the financial advisor.}

The only task of the financial advisors in this experiment is to forecast the stock
market index in each time period as accurate as possible. The stock price has to be
predicted two time periods ahead. At the beginning of the experiment begins, you
have to predict the stock price in the first two periods. It is very likely that the
stock price will be between 0 and 100 in the first two periods. After all participants
have given their predictions for the first two periods, the stock market price for the
first period will be revealed and based upon your forecasting error your earnings for
period 1 will be given. After that you have to give your prediction for the stock
market index in the third period. After all participants have given their predictions
for period 3, the stock market index in the second period will be revealed and, based
upon your forecasting error your earnings for period 2 will be given. This process
continues for 51 time periods.

The available information for forecasting the stock price in period t consists of:

\begin{itemize}
    \item all past prices up to period t - 2, and
    \item past predictions up to period t - 1, and
    \item total earnings up to period t - 2
\end{itemize}

\textbf{Information about the stock market.}

The stock market price is determined by equilibrium between demand and supply
of stocks. The stock market price in period t will be that price for which aggregate
demand equals supply. The supply of stocks is fixed during the experiment. The
demand for stocks is determined by the aggregate demand of a number of large
pension funds active. Each pension fund is advised by a participant of the experiment.

\textbf{Information about the investment strategies of the pension funds.}

The precise investment strategy of the pension fund that you are advising and the
1
investment strategies of the other pension funds are unknown. The bank account of
the risk free investment pays a fixed interest rate of 5\% per time period. The holder of
the stock receives a dividend payment in each time period. These dividend payments
are uncertain however and vary over time. Economic experts of the pension funds
have computed that the average dividend payments are 3 guilder per time period. The
return of the stock market per time period is uncertain and depends upon (unknown)
dividend payments as well as upon price changes of the stock. As the financial advisor
of a pension fund you are not asked to forecast dividends, but you are only asked to
forecast the price of the stock in each time period. Based upon your stock market
price forecast, your pension fund will make an optimal investment decision. The
higher your price forecast the larger will be the fraction of money invested by your
pension fund in the stock market, so the larger will be their demand for stocks.

\textbf{Earnings.}

Earnings will depend upon forecasting accuracy only. The better you predict the
stock market price in each period, the higher your aggregate earnings.
Earnings will be according to the following earnings table.

\begin{table}[h]
\centering
\caption{Earnings (Payoff) Table}
\label{tab:earnings}
\small
\begin{tabular}{cc cc cc cc cc}
\hline
Error & Points & Error & Points & Error & Points & Error & Points & Error & Points \\
\hline
0.10 & 1300 & 1.50 & 1240 & 3.00 & 1061 & 4.40 & 786 & 5.80 & 408 \\
0.15 & 1299 & 1.55 & 1236 & 3.05 & 1053 & 4.45 & 775 & 5.85 & 392 \\
0.20 & 1299 & 1.60 & 1232 & 3.10 & 1045 & 4.50 & 763 & 5.90 & 376 \\
0.25 & 1298 & 1.65 & 1228 & 3.15 & 1037 & 4.55 & 751 & 5.95 & 361 \\
0.30 & 1298 & 1.70 & 1223 & 3.20 & 1028 & 4.60 & 739 & 6.00 & 345 \\
0.35 & 1297 & 1.75 & 1219 & 3.25 & 1020 & 4.65 & 726 & 6.05 & 329 \\
0.40 & 1296 & 1.80 & 1214 & 3.30 & 1011 & 4.70 & 714 & 6.10 & 313 \\
0.45 & 1295 & 1.85 & 1209 & 3.35 & 1002 & 4.75 & 701 & 6.15 & 297 \\
0.50 & 1293 & 1.90 & 1204 & 3.40 & 993  & 4.80 & 689 & 6.20 & 280 \\
0.55 & 1292 & 1.95 & 1199 & 3.45 & 984  & 4.85 & 676 & 6.25 & 264 \\
0.60 & 1290 & 2.00 & 1194 & 3.50 & 975  & 4.90 & 663 & 6.30 & 247 \\
0.65 & 1289 & 2.05 & 1189 & 3.55 & 966  & 4.95 & 650 & 6.35 & 230 \\
0.70 & 1287 & 2.10 & 1183 & 3.60 & 956  & 5.00 & 637 & 6.40 & 213 \\
0.75 & 1285 & 2.15 & 1177 & 3.65 & 947  & 5.05 & 623 & 6.45 & 196 \\
0.80 & 1283 & 2.20 & 1172 & 3.70 & 937  & 5.10 & 610 & 6.50 & 179 \\
0.85 & 1281 & 2.25 & 1166 & 3.75 & 927  & 5.15 & 596 & 6.55 & 162 \\
0.90 & 1279 & 2.30 & 1160 & 3.80 & 917  & 5.20 & 583 & 6.60 & 144 \\
0.95 & 1276 & 2.35 & 1153 & 3.85 & 907  & 5.25 & 569 & 6.65 & 127 \\
1.00 & 1273 & 2.40 & 1147 & 3.90 & 896  & 5.30 & 555 & 6.70 & 109 \\
1.05 & 1271 & 2.45 & 1141 & 3.95 & 886  & 5.35 & 541 & 6.75 & 91 \\
1.10 & 1268 & 2.60 & 1121 & 4.00 & 876  & 5.40 & 526 & 6.80 & 73 \\
1.15 & 1265 & 2.65 & 1114 & 4.05 & 865  & 5.45 & 512 & 6.85 & 55 \\
1.20 & 1262 & 2.70 & 1107 & 4.10 & 854  & 5.50 & 497 & 6.90 & 37 \\
1.25 & 1259 & 2.75 & 1099 & 4.15 & 843  & 5.55 & 483 & 6.95 & 19 \\
1.30 & 1255 & 2.80 & 1092 & 4.20 & 832  & 5.60 & 468 & $\geq 7.00$ & 0 \\
1.35 & 1252 & 2.85 & 1085 & 4.25 & 821  & 5.65 & 453 &  &  \\
1.40 & 1248 & 2.90 & 1077 & 4.30 & 809  & 5.70 & 438 &  &  \\
1.45 & 1244 & 2.95 & 1069 & 4.35 & 798  & 5.75 & 423 &  &  \\
\hline
\end{tabular}
\vspace{0.5em}

\footnotesize{Note: 1300 points equal 1 Dutch guilder.}
\end{table}

\newpage

\subsection{LLM experiment prompts}
\label{app:LLMprompts}

\textbf{System prompt}

\begin{promptbox}
General information: You are a financial advisor to a pension fund that wants to optimally invest a large amount of money. The pension fund has two investment options: a risk free investment and a risky investment. The risk free investment is putting all money on a bank account paying a fixed and known interest rate. The alternative risky investment is an investment in the stock market with uncertain return. In each time period the pension fund has to decide which fraction of its money to put on the bank account and which fraction of the money to spend on buying stocks. In order to make an optimal investment decision the pension fund needs an accurate prediction of the price of the stock. As their financial advisor, you have to predict the stock market price during {} subsequent time periods. Your earnings during the experiment depend upon your forecasting accuracy. The smaller your forecasting errors in each period, the higher your total earnings.

Forecasting task of the financial advisor: The only task of the financial advisors in this experiment is to forecast the stock market index in each time period as accurate as possible. The stock price has to be predicted two time periods ahead. At the beginning of the experiment begins, you have to predict the stock price in the first two periods. It is very likely that the stock price will be between 0 and 100 in the first two periods. After all participants have given their predictions for the first two periods, the stock market price for the first period will be revealed and based upon your forecasting error your earnings for period 1 will be given. After that you have to give your prediction for the stock market index in the third period. After all participants have given their predictions for period 3, the stock market index in the second period will be revealed and, based upon your forecasting error your earnings for period 2 will be given. This process continues for {} time periods.
The available information at period t for forecasting the stock price in period t consists of:
- all past prices up to period t-2, and
- past predictions up to period t-1, and
- total earnings up to period t-2

Information about the stock market: The stock market price is determined by equilibrium between demand and supply of stocks. The stock market price in period t will be that price for which aggregate demand equals supply. The supply of stocks is fixed during the experiment. The demand for stocks is determined by the aggregate demand of a number of large pension funds active. Each pension fund is advised by a participant of the experiment.

Information about the investment strategies of the pension funds: The precise investment strategy of the pension fund that you are advising and the investment strategies of the other pension funds are unknown. The bank account of the risk free investment pays a fixed interest rate of 5\% per time period. The holder of the stock receives a dividend payment in each time period. These dividend payments are uncertain however and vary over time. Economic experts of the pension funds have computed that the average dividend payments are 3 guilder per time period. The return of the stock market per time period is uncertain and depends upon (unknown) dividend payments as well as upon price changes of the stock. As the financial advisor of a pension fund you are not asked to forecast dividends, but you are only asked to forecast the price of the stock in each time period. Based upon your stock market price forecast, your pension fund will make an optimal investment decision. The higher your price forecast the larger will be the fraction of money invested by your pension fund in the stock market, so the larger will be their demand for stocks.

Earnings: earnings will depend upon forecasting accuracy only. The better you predict the stock market price in each period, the higher your aggregate earnings. Earnings will be according to the following earnings table.

| error | points |
|-------|--------|
| 0.1   | 1300   |
| 0.15  | 1299   |
| 0.2   | 1299   |
| 0.25  | 1298   |
| 0.3   | 1298   |
| 0.35  | 1297   |
| 0.4   | 1296   |
| 0.45  | 1295   |
| 0.5   | 1293   |
| 0.55  | 1292   |
| 0.6   | 1290   |
| 0.65  | 1289   |
| 0.7   | 1287   |
| 0.75  | 1285   |
| 0.8   | 1283   |
| 0.85  | 1281   |
| 0.9   | 1279   |
| 0.95  | 1276   |
| 1     | 1273   |
| 1.05  | 1271   |
| 1.1   | 1268   |
| 1.15  | 1265   |
| 1.2   | 1262   |
| 1.25  | 1259   |
| 1.3   | 1255   |
| 1.35  | 1252   |
| 1.4   | 1248   |
| 1.45  | 1244   |
| 1.5   | 1240   |
| 1.55  | 1236   |
| 1.6   | 1232   |
| 1.65  | 1228   |
| 1.7   | 1223   |
| 1.75  | 1219   |
| 1.8   | 1214   |
| 1.85  | 1209   |
| 1.9   | 1204   |
| 1.95  | 1199   |
| 2     | 1194   |
| 2.05  | 1189   |
| 2.1   | 1183   |
| 2.15  | 1177   |
| 2.2   | 1172   |
| 2.25  | 1166   |
| 2.3   | 1160   |
| 2.35  | 1153   |
| 2.4   | 1147   |
| 2.45  | 1141   |
| 2.6   | 1121   |
| 2.65  | 1114   |
| 2.7   | 1107   |
| 2.75  | 1099   |
| 2.8   | 1092   |
| 2.85  | 1085   |
| 2.9   | 1077   |
| 2.95  | 1069   |
| 3     | 1061   |
| 3.05  | 1053   |
| 3.1   | 1045   |
| 3.15  | 1037   |
| 3.2   | 1028   |
| 3.25  | 1020   |
| 3.3   | 1011   |
| 3.35  | 1002   |
| 3.4   | 993    |
| 3.45  | 984    |
| 3.5   | 975    |
| 3.55  | 966    |
| 3.6   | 956    |
| 3.65  | 947    |
| 3.7   | 937    |
| 3.75  | 927    |
| 3.8   | 917    |
| 3.85  | 907    |
| 3.9   | 896    |
| 3.95  | 886    |
| 4     | 876    |
| 4.05  | 865    |
| 4.1   | 854    |
| 4.15  | 843    |
| 4.2   | 832    |
| 4.25  | 821    |
| 4.3   | 809    |
| 4.35  | 798    |
| 4.4   | 786    |
| 4.45  | 775    |
| 4.5   | 763    |
| 4.55  | 751    |
| 4.6   | 739    |
| 4.65  | 726    |
| 4.7   | 714    |
| 4.75  | 701    |
| 4.8   | 689    |
| 4.85  | 676    |
| 4.9   | 663    |
| 4.95  | 650    |
| 5     | 637    |
| 5.05  | 623    |
| 5.1   | 610    |
| 5.15  | 596    |
| 5.2   | 583    |
| 5.25  | 569    |
| 5.3   | 555    |
| 5.35  | 541    |
| 5.4   | 526    |
| 5.45  | 512    |
| 5.5   | 497    |
| 5.55  | 483    |
| 5.6   | 468    |
| 5.65  | 453    |
| 5.7   | 438    |
| 5.75  | 423    |
| 5.8   | 408    |
| 5.85  | 392    |
| 5.9   | 376    |
| 5.95  | 361    |
| 6     | 345    |
| 6.05  | 329    |
| 6.1   | 313    |
| 6.15  | 297    |
| 6.2   | 280    |
| 6.25  | 264    |
| 6.3   | 247    |
| 6.35  | 230    |
| 6.4   | 213    |
| 6.45  | 196    |
| 6.5   | 179    |
| 6.55  | 162    |
| 6.6   | 144    |
| 6.65  | 127    |
| 6.7   | 109    |
| 6.75  | 91     |
| 6.8   | 73     |
| 6.85  | 55     |
| 6.9   | 37     |
| 6.95  | 19     |
| >= 7  | 0      |

Response format:  Your response should be exclusively in JSON format with keys: 'reasoning' where you explain your rationale and method for prediction, and 'predictedValue', the numeric value of your predicted market price. Nothing outside the JSON format should be written. Only for the first time step should there be three keys: 'reasoning', 'predictedValue1', and 'predictedValue2'.

\end{promptbox}

\textbf{Initial user prompt}

\begin{promptbox}
This is the first time step, give an initial price prediction for the first two periods.  It is very likely that the stock price will be between 0 and 100 in the first two periods.

Your response should be in the format:

\{"reasoning": "your reasoning here", "predictedValue1": xx.xx, "predictedValue2": xx.xx\}
\end{promptbox}

\textbf{Subsequent user prompt}

\begin{promptbox}
Current time step: \{\}; below is a markdown table showing historical market prices and your predictions: 
                                    
\{\}

Total earnings up to time \{\}: \{\}. Earnings at last time step: \{\}.

Make your prediction for what the price will be in time period \{\}. 

Your response should be in the format:

\{"reasoning": "your reasoning here", "predictedValue": xx.xx\}
\end{promptbox}

\textbf{Price cap message}

\begin{promptbox}
Predictions above 1000 or below 0 are not accepted please submit a prediction between 0 and 1000.
\end{promptbox}

\subsection{Bubble measures definitions}
\label{app:bubblemeasures}

Let $p_t$ denote the realised market price at time $t$, $T$ be the total number of periods, $p^f$ the fundamental value (where $p^f = 60$ for our experiment) and $\bar{p}$ be the mean realised price. The following measures are used to quantify price deviations and market volatility in the main body of the text:

\begin{enumerate}
    \item \textbf{Relative Deviation (RD):} Measures the mean overpricing relative to the fundamental price.
    \begin{equation}
        RD = \frac{1}{T} \sum_{t=1}^{T} \frac{p_t - p^f}{p^f}
    \end{equation}

    \item \textbf{Maximum Relative Deviation ($RDMAX$):} Measures the peak amplitude (the highest relative bubble) reached during the session.
    \begin{equation}
        RDMAX = \max_t \left( \frac{p_t - p^f}{p^f} \right)
    \end{equation}

    \item \textbf{Price Variability:} To measure the variability of price action, we employ the standard deviation and the interquartile range (IQR) of the price series:
    \begin{equation}
        \sigma_p = \sqrt{\frac{1}{T-1} \sum_{t=1}^{T} (p_t - \bar{p})^2}
    \end{equation}
    \begin{equation}
        IQR = Q_3(p_t) - Q_1(p_t)
    \end{equation}
\end{enumerate}

We also introduce the below measures for robustness, finding that our results remain valid across several different measures:

\begin{enumerate}
    \item \textbf{Relative Absolute Deviation (RAD):} Measures the mean magnitude of absolute mispricing.
    \begin{equation}
        RAD = \frac{1}{T} \sum_{t=1}^{T} \frac{|p_t - p^f|}{p^f}
    \end{equation}

    \item \textbf{Geometric Deviation (GD):} A numeraire-independent equivalent of RD from \citet{POWELL201656}.
    \begin{equation}
        GD = \exp\left( \frac{1}{T} \sum_{t=1}^{T} \ln \left( \frac{p_t}{p^f} \right) \right) - 1
    \end{equation}

    \item \textbf{Geometric Absolute Deviation (GAD):} A numeraire-independent equivalent of RAD.
    \begin{equation}
        GAD = \exp\left( \frac{1}{T} \sum_{t=1}^{T} \left| \ln \left( \frac{p_t}{p^f} \right) \right| \right) - 1
    \end{equation}
\end{enumerate}

\subsection{Robustness of bubble indicators}
\label{app:robustnesschecks}

There is no universally accepted definition of an experimental price bubble. We define our own measure, $P_{bubble}$, since it is convenient for parts of our analysis to have a binary indicator showing that a bubble has formed. We define a bubble as forming if the market price remains above five times the fundamental price (300) for three or more consecutive periods. To test the robustness of this binary indicator, we compare it to the bubble measure used in \citet{kopanyi2021experience, kopanyi2024role}, which we denote $P_{bubble}^{mean}$. This classifies a price series as containing a bubble if the mean price is above twice the fundamental price (120 in this experiment). 

Both of these indicators have variable hyperparameters: $P_{bubble}$ has the price threshold and duration threshold while $P_{bubble}^{mean}$ has only the price threshold. To test robustness, we classify the results across 15 models, with a total of 188 experiments of which roughly 30\% are found to include a bubble. For $P_{bubble}$ we consider five thresholds (180, 240, 300, 360, 420) and three lengths of time (3, 5, 7), giving 15 classifications.  For $P_{bubble}^{mean}$, we consider four thresholds (90, 120, 150, 180). For each experimental run, this gives us 19 classifications. We then compute Cohen's kappa between each classification pair to measure agreement between the measures. Cohen's kappa ranges from -1 to 1, with higher values signalling stronger agreement. We find that all Cohen kappa scores are above 0.85, indicating very strong agreement between the two methods and across parameter values.

\newpage
\section{Results}

\subsection{Full results table with added measures}
\label{app:results:measures}

\begin{table}[htbp]
\centering
\setlength{\tabcolsep}{3pt}
\renewcommand{\arraystretch}{0.9}
\begin{tabular}{lrrrrrrrr}
\toprule
 &  & \multicolumn{5}{c}{Overpricing} & \multicolumn{2}{c}{Vol.} \\
\cmidrule(lr){3-7} \cmidrule(lr){8-9}
Model & Repeats & RD & RAD & GD & GAD & RDMAX & IQR pr & Std pr \\
\midrule
o3-mini & 6 & 6.17 & 6.19 & 4.26 & 4.37 & 14.21 & 457.91 & 284.41 \\
OLMO-Think & 17 & 6.08 & 6.39 & 2.54 & 4.90 & 14.90 & 666.37 & 353.74 \\
\rowcolor{MySoftGreen} RE (constant = 400) &  & 3.62 & 3.62 & 1.29 & 1.29 & 15.28 & 504.44 & 318.29 \\
Qwen3-14B & 20 & 3.25 & 3.45 & 1.14 & 1.85 & 14.32 & 279.40 & 301.37 \\
\rowcolor{MySoftGreen} RE (constant = 575) &  & 2.53 & 2.53 & 0.68 & 0.68 & 15.61 & 0.00 & 310.57 \\
\rowcolor{MyLightBlue} Humans & 6 & 2.40 & 2.60 & 0.99 & 1.68 & 12.16 & 213.77 & 215.63 \\
Qwen3-32B & 15 & 2.12 & 2.24 & 0.86 & 1.61 & 6.13 & 228.52 & 143.85 \\
o3 & 6 & 1.42 & 1.47 & 0.55 & 0.66 & 12.50 & 62.50 & 186.81 \\
GPT-4o mini & 6 & 0.26 & 0.28 & 0.25 & 0.27 & 0.58 & 19.38 & 12.20 \\
OLMO-Instruct & 20 & 0.14 & 0.18 & 0.13 & 0.18 & 0.28 & 10.51 & 7.64 \\
Qwen2.5-7B & 20 & 0.08 & 0.14 & 0.07 & 0.14 & 0.30 & 13.10 & 8.09 \\
R1 - Qwen & 17 & 0.08 & 0.12 & 0.08 & 0.12 & 0.22 & 8.96 & 6.07 \\
R1 - Llama & 19 & 0.02 & 0.14 & 0.01 & 0.15 & 0.12 & 3.70 & 3.78 \\
GPT-5 mini & 6 & 0.01 & 0.01 & 0.00 & 0.01 & 0.02 & 0.50 & 0.96 \\
Gemini-2.5-Flash & 6 & -0.00 & 0.00 & -0.00 & 0.00 & 0.00 & 0.00 & 0.04 \\
Gemini-3-Flash & 6 & 0.00 & 0.00 & 0.00 & 0.00 & 0.00 & 0.00 & 0.00 \\
GPT-4.1 & 6 & 0.00 & 0.04 & 0.00 & 0.04 & 0.10 & 3.42 & 3.26 \\
\rowcolor{MySoftGreen} RE (constant = 0) &  & 0.00 & 0.00 & 0.00 & 0.00 & 0.00 & 0.00 & 0.00 \\
Gemma & 18 & -0.04 & 0.06 & -0.04 & 0.07 & 0.01 & 3.93 & 3.14 \\
\bottomrule
\end{tabular}
\caption{\textbf{Quantitative analysis shows a wide range of results, robust across different measures.} Statistics are averaged across independent experimental results. Rpts is the number of independent repeats.  Relative Deviation (RD) and Geometric Deviation (GD) are a measure of over-pricing measured as a multiple of the fundamental value. Relative Absolute Deviation (RAD) and Geometric Absolute Deviation (GAD) are measures of mispricing as a multiple of the fundamental value. RDMAX is the maximum overpricing. IQR pr is the inter-quartile range of the prices and Std pr is the standard deviation of the prices. Models are sorted by descending RD. Human results from \citet{hommes2008expectations} are highlighted in blue, while rational expectations (RE) solutions are highlighted in green.}
\label{app:tab:bubbchar}
\end{table}

\subsection{Robustness to changing temperature and memory}
\label{app:results:robustsinglellm}

We test robustness of our results across three temperatures (0.3,0.7,1.0) and three memory parameters (0,2,4). We do this for representative models of the three categories of LLM: OLMO-Instruct (for no bubbles but not rational expectations), Qwen3-14B (for speculative bubbles) and Gemini-3-Flash (for models well described by rational expectations). The results are given in Table \ref{app:tab:singlellmrobustness}.

\begin{table}[htbp]
\centering
\setlength{\tabcolsep}{3pt}
\renewcommand{\arraystretch}{0.9}
\begin{tabular}{lllrrrrrrr}
\toprule
 & \multicolumn{2}{c}{Settings} & & \multicolumn{2}{c}{Overpricing} & \multicolumn{1}{c}{Vol.} & & \multicolumn{2}{c}{Non-REH} \\
\cmidrule(lr){2-3} \cmidrule(lr){5-6} \cmidrule(lr){7-7} \cmidrule(lr){9-10} 
Model & Temp. & Memory & Repeats & RD & RDMAX & IQR pr & $P_{bubble}$ & SPEC & BIAS \\
\midrule
Qwen3-14B & 0.70 & 2 & 5 & 4.13 & 14.89 & 426.46 & 1.00 & 1.00 & 0.73 \\
Qwen3-14B & 0.30 & 2 & 4 & 3.83 & 14.85 & 456.59 & 1.00 & 1.00 & 0.67 \\
Qwen3-14B & 0.70 & 4 & 5 & 3.58 & 14.58 & 357.74 & 1.00 & 0.80 & 0.73 \\
Qwen3-14B & 0.30 & 0 & 5 & 3.54 & 14.89 & 392.03 & 1.00 & 1.00 & 0.43 \\
Qwen3-14B & 0.70 & 0 & 4 & 3.48 & 14.89 & 367.52 & 1.00 & 0.75 & 0.54 \\
Qwen3-14B & 1 & 4 & 5 & 3.35 & 14.82 & 316.42 & 1.00 & 1.00 & 0.77 \\
Qwen3-14B & 1 & 2 & 20 & 3.25 & 14.32 & 279.40 & 1.00 & 0.95 & 0.75 \\
Qwen3-14B & 0.30 & 4 & 5 & 3.22 & 9.54 & 371.01 & 0.60 & 1.00 & 0.70 \\
Qwen3-14B & 1 & 0 & 5 & 3.21 & 14.92 & 293.95 & 1.00 & 0.80 & 0.50 \\
OLMO-Instruct & 1 & 4 & 5 & 0.19 & 0.48 & 17.93 & 0.00 & - & 0.93 \\
OLMO-Instruct & 0.70 & 2 & 5 & 0.18 & 0.46 & 25.73 & 0.00 & - & 0.93 \\
OLMO-Instruct & 1 & 0 & 5 & 0.18 & 0.33 & 12.15 & 0.00 & - & 0.93 \\
OLMO-Instruct & 0.70 & 0 & 5 & 0.14 & 0.29 & 15.40 & 0.00 & - & 1.00 \\
OLMO-Instruct & 1 & 2 & 20 & 0.14 & 0.28 & 10.51 & 0.00 & - & 0.95 \\
OLMO-Instruct & 0.30 & 0 & 5 & 0.13 & 0.34 & 15.11 & 0.00 & - & 1.00 \\
OLMO-Instruct & 0.30 & 4 & 5 & 0.13 & 0.43 & 15.46 & 0.00 & - & 0.97 \\
OLMO-Instruct & 0.70 & 4 & 5 & 0.11 & 0.36 & 15.27 & 0.00 & - & 1.00 \\
OLMO-Instruct & 0.30 & 2 & 5 & 0.05 & 0.18 & 9.04 & 0.00 & - & 0.90 \\
Gemini-3-Flash & 0.30 & 2 & 1 & 0.00 & 0.00 & 0.00 & 0.00 & - & 0.00 \\
Gemini-3-Flash & 0.30 & 0 & 1 & 0.00 & 0.00 & 0.00 & 0.00 & - & 0.00 \\
Gemini-3-Flash & 0.70 & 2 & 1 & 0.00 & 0.00 & 0.00 & 0.00 & - & 0.00 \\
Gemini-3-Flash & 0.70 & 0 & 1 & 0.00 & 0.00 & 0.00 & 0.00 & - & 0.00 \\
Gemini-3-Flash & 0.30 & 4 & 1 & 0.00 & 0.00 & 0.00 & 0.00 & - & 0.00 \\
Gemini-3-Flash & 0.70 & 4 & 1 & 0.00 & 0.00 & 0.00 & 0.00 & - & 0.00 \\
Gemini-3-Flash & 1 & 4 & 1 & 0.00 & 0.00 & 0.00 & 0.00 & - & 0.00 \\
Gemini-3-Flash & 1 & 2 & 6 & 0.00 & 0.00 & 0.00 & 0.00 & - & 0.00 \\
Gemini-3-Flash & 1 & 0 & 1 & 0.00 & 0.00 & 0.00 & 0.00 & - & 0.00 \\
\bottomrule
\end{tabular}
\caption{\textbf{Key results are robust across memory and temperature values.} Statistics are averaged across independent experimental results. Temp is the temperature of the LLM. Rpts is the number of independent repeats.  Relative Deviation (RD) is a measure of over-pricing measured as a multiple of the fundamental value, while RDMAX is the maximum overpricing. IQR pr is the inter-quartile range of the prices while $P_{bubble}$ is the proportion of runs which formed a bubble, according to the metric introduced in this paper.  The proportion of experiments with provably speculative bubbles is SPEC and BIAS is the proportion of biased agents.  Models are sorted by descending RD.}
\label{app:tab:singlellmrobustness}
\end{table}

We find that our key results are robust to varying memory and temperature. Table \ref{app:tab:singlellmrobustness} shows that Gemini-3-Flash continues to predict only the fundamental value across settings, while OLMO-Instruct never forms bubbles but has almost all agents as provably biased. This means that OLMO-Instruct is not well captured by the rational expectations hypothesis, regardless of parameter settings. Finally Qwen3-14B always makes a bubble, with the only exception two runs (out of 58) with 0.3 temperature and a memory of 4. Qwen3-14B's bubbles remain provably speculative and lead to far higher volatility and mispricing than results of OLMO-Instruct or Gemini-3-Flash.

\subsection{Analysis of which LLMs form bubbles}
\label{app:results:whichLLMs}

We find that in order for LLMs to form speculative bubbles, they must be of a middling capability and sufficiently aggressive in their extrapolations of trends. We focus on mathematical and scientific capabilities, with a heavier weight on reasoning than pure memorised knowledge. We do this by creating a capability index using an equally weighted average of three tests: Humanity's Last Exam, AIME 25 and SciCode. The results of these three tests are taken from independent testing from \citet{artificialanalysis_2026_aimodel}. Each score is then min-max normalised to lie between 0 and 1 and the mean is taken to give a final capability index. 

Aggressiveness of trend extrapolation is tested by putting each LLM agent into a simulated and identical snapshot of the \citet{hommes2008expectations} experiment we use in this paper. Given the historical prices, historical predictions and all prompts given to each LLM agent are identical, any differences in price prediction is purely from differences in the LLMs.  The price sequence given to the LLMs shows a gentle convex curve higher (as in Table \ref{app:tab:testmarketprices}), which could feasibly be extrapolated in many different ways including linearly or non-linearly. Agents are asked to predict the market price for time step 8. This is repeated five times and we take the mean price prediction as a measure of how aggressively the trend is extrapolated.

\begin{table}[h!]
\centering
\begin{tabular}{|c|c|c|}
\hline
\textbf{Time Step} & \textbf{Market Price} & \textbf{Your Prediction} \\
\hline
1 & 53.65 & 50.0 \\
2 & 55.18 & 50.0 \\
3 & 57.52 & 51.83 \\
4 & 60.47 & 56.71 \\
5 & 63.49 & 61.26 \\
6 & 67.47 & 64.69 \\
7 & N/A & 68.06 \\
\hline
\end{tabular}
\label{app:tab:testmarketprices}
\caption{Market prices and predictions given to LLMs to test aggressiveness of extrapolation.}
\end{table}

\begin{figure}[tbp]
    \centering
    \includegraphics[width=\linewidth]{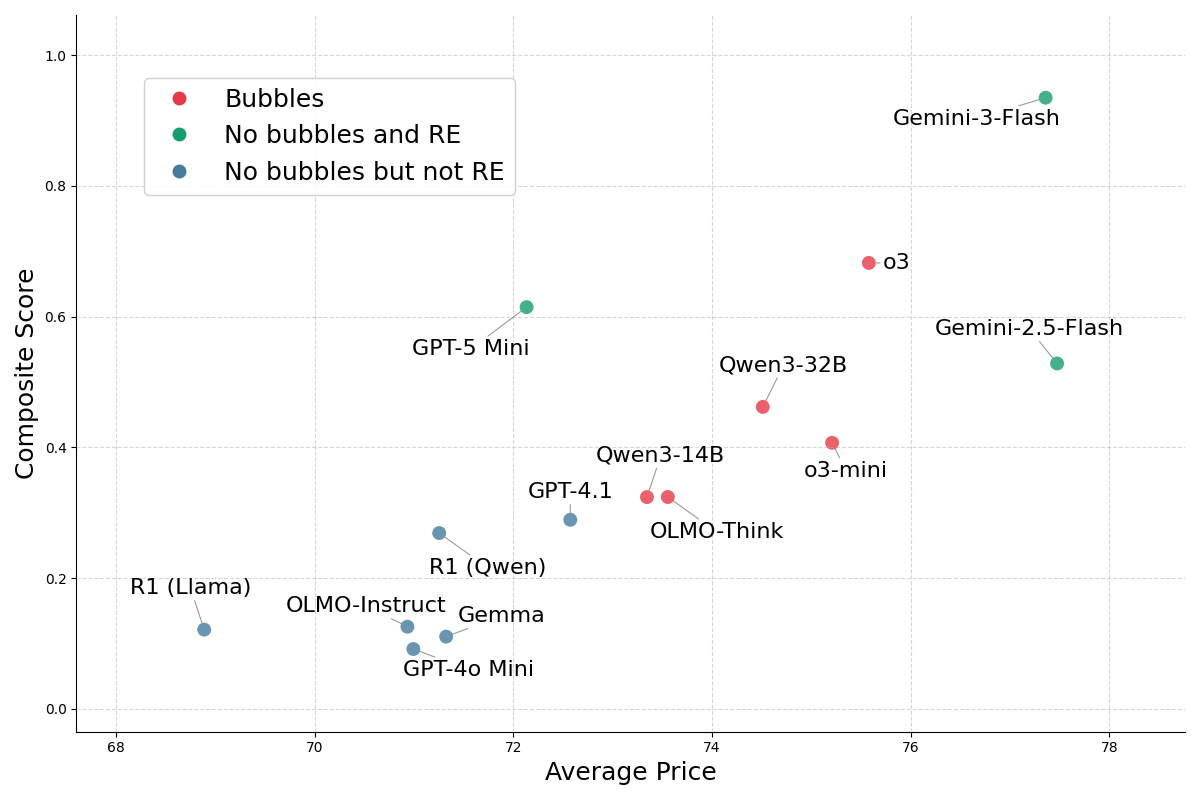}
    \caption{\textbf{Middling capability and aggressive price prediction are good predictors of bubble formation.} Composite score refers to the capability index introduced in this paper. Average price is the mean price prediction from the extrapolation aggressiveness test.}
    \label{app:fig:scorevsprice}
\end{figure}

We show the classifications for each LLM in Figure \ref{app:fig:scorevsprice}. We find that low capability and low aggressiveness translates to markets where no speculative bubbles form but the results are not consistent with the rational expectations hypothesis. Meanwhile, high capability is generally associated with predicting the fundamental price (aligning with the RE fundamental solution), regardless of how aggressively the LLM extrapolates prices. The middle ground, LLMs with middling capabilities but sufficiently aggressive price extrapolation, are those which form large price bubbles.

\subsection{How are price bubbles formed?}
\label{app:results:whybubbles}

We also consider what characterises the predictions made by agents when they form price bubbles. 

\subsubsection{Methods}

We analyse how bubbles are formed by using a classifier on the justifications outputted by each agent for their price prediction. For the non-linear extrapolation, we focus on periods after time step 3 (so that non-linear extrapolation makes mathematical sense) and before either the first bubble peak (as defined in Section \ref{subsec:bubble_measures}) or time step 12. This is done so that the effects of the 1000 price cap do not distort results, and so that the number of time steps considered are roughly equal across experiments with and without bubbles. The fundamental classification uses the same time periods but also includes the first three time steps. We use an LLM (Gemma3-27B) to make these classifications, using the prompts in Section \ref{app:classifierprompts}. 

The Gemma classifications are validated by comparing against both human and GPT-5 Mini classifications using the same prompt. For validation against human labels, 200 justifications were randomly selected (across time steps, models and agents) and each justification was labelled as containing non-linear reasoning and/or anchoring to a fundamental value. Comparison of human and Gemma classifications gave a Cohen's Kappa score of 0.7 and 0.91 for the non-linear and fundamental classification respectively. A similar procedure was carried out using GPT-5 Mini to perform both classifications over 500 randomly selected justifications. Comparison of Gemma and GPT-5 Mini classifications give a Cohen's Kappa of 0.59 for the non-linear classification, and 0.75 for the fundamental classification.  These high values show that our classification method is robust across two alternatives. 

\subsubsection{LLM classifier prompts}
\label{app:classifierprompts}

Prompt for LLM classification of whether an agent used non-linear extrapolation in their price prediction:

\begin{promptbox}
Analyse the following response from an agent trying to predict the price of an asset in a market experiment. Identify if the agent is clearly and explicitly using non-linear extrapolation to make their prediction, and explain your reasoning.  Some examples of non-linear reasoning are:

- Uses a quadratic or higher order polynomial
- Uses an exponential model
- Reasons about the returns of the price, or percentage change in price, rather than about absolute price changes
- Reasons about the rate of change of the differences in prices.

Your response should be in exclusively in JSON format, with the structure below:

\{{"reasoning": "your reasoning here", "Non-linear extrapolation": 1 if non-linear extrapolation is used, 0 otherwise}\}

Here is the text to analyse:

\{\}
\end{promptbox}

Prompt for LLM classification of whether an agent used a fundamental valuation in their price prediction:

\begin{promptbox}
Analyse the following response from an agent trying to predict the price of an asset in a market experiment. Identify if the agent is clearly and explicitly anchoring their prediction around what they consider is a fundamental asset price.  This fundamental price can be calculated in any way, such as dividend discounting or any other method. For reference, the correct fundamental price in this experiment is 60 but the agent could be anchoring to a different, incorrect, fundamental value.

Your response should be in exclusively in JSON format, with the structure below:

\{{"reasoning": "your reasoning here", "Fundamental": 1 if fundamental price is used, 0 otherwise}\}

Here is the text to analyse:

\{\}
\end{promptbox}

\subsubsection{Classification results} 

\begin{figure}[htbp]
    \centering
    \includegraphics[width=\linewidth]{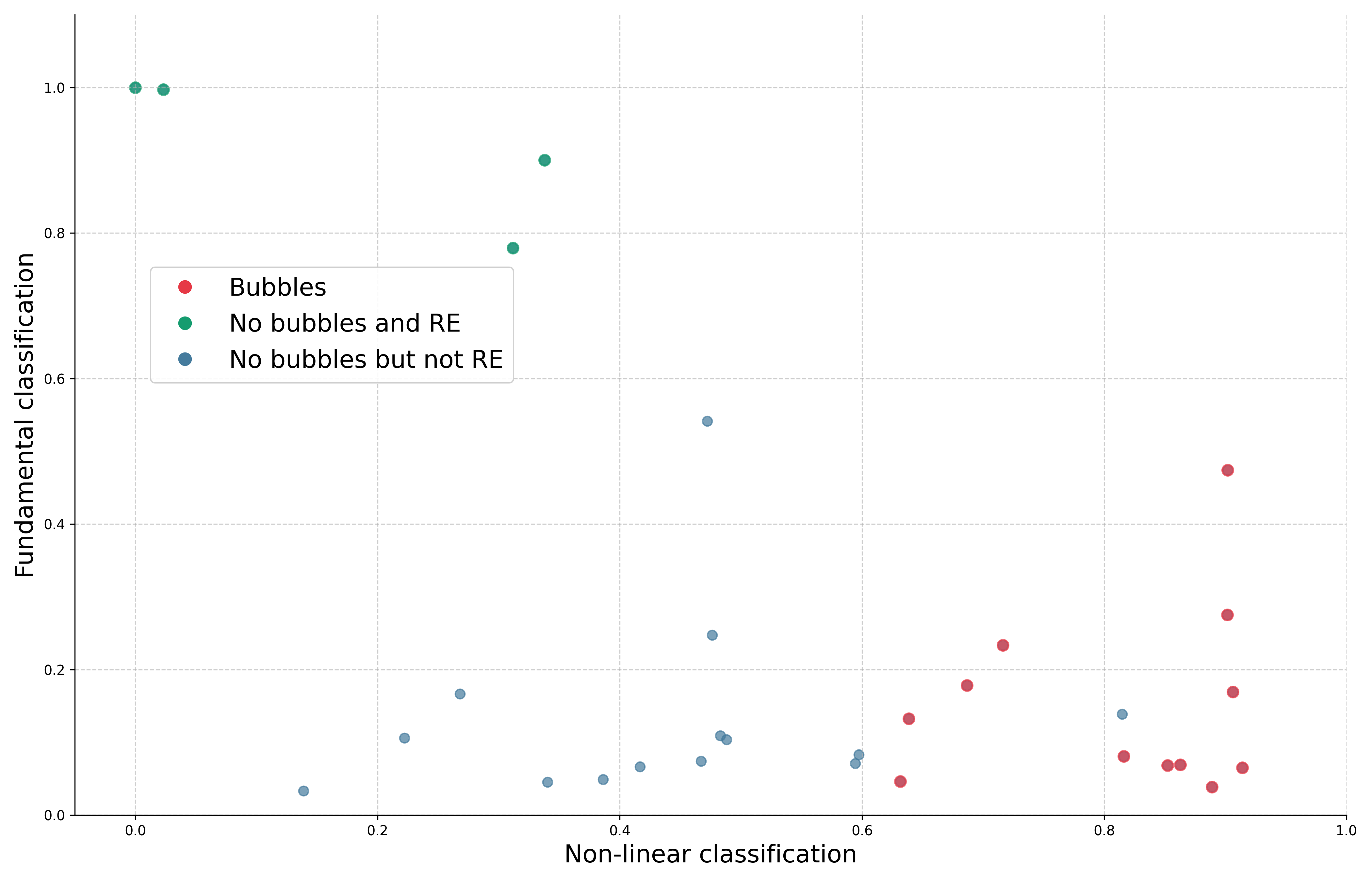}
    \caption{\textbf{Bubble experiments have on average a high proportion of non-linear extrapolation methods used and a low proportion of price predictions anchored to the fundamental value.} Proportions of predictions using non-linear extrapolation methods and anchoring to the fundamental value, averaged across experimental runs conditional on whether a bubble formed or not. For example, one point is the mean over Qwen3-32B runs which form bubbles, and another point is the mean over Qwen3-32B runs which do not form bubbles.}
    \label{app:res:justificationclassification}
\end{figure}

Both the non-linear extrapolation and fundamental value classification tasks are run on all experimental repeats of single-LLM markets and mixed markets, with the results shown in Figure \ref{app:res:justificationclassification}.

We find that experiments in which bubbles form have high proportions of non-linear extrapolation methods used and low anchoring to the fundamental value. This suggests that bubble formation is associated with high numbers of aggressive trend-followers and low numbers of fundamentalists. There is one outlier in the bottom right corner which appears to have high proportions of non-linear methods and low references to the fundamental value, and yet does not form bubbles. These results are from GPT-4.1 and are because this LLM tends to use non-linear methods but tempers its predictions on the way up so that the initial phase of the experiment is roughly linear extrapolation in practice. This means that full bubbles never form, as the \enquote{pull} of the fundamental price wins over and the realised price fluctuates around the fundamental value.  

\subsection{Causal analysis}
\label{app:results:causalanalysis}

We also explore some causal questions. The first we look at is the causal effect of reasoning capabilities on bubble formation.

\subsubsection{Causal impact of reasoning on bubble formation}

We explore this question using Qwen3-14B's reasoning toggle. This switch turns on and off the \enquote{thinking} tokens, tokens that are outputted in a separate block before the final answer is given, without changing the model. We can therefore explore the impact this reasoning has while keeping everything else the same, and therefore make causal claims about the impact of these reasoning tokens on the propensity of Qwen3-14B to form bubbles. 

We find that Qwen3-14B without reasoning produces zero speculative bubbles (over 40 independent repeats), while Qwen3-14B with reasoning always produces speculative bubbles (over 20 independent repeats). We also find that Qwen3-14B with reasoning has higher mispricing and volatility compared to Qwen3-14B without reasoning (significant at the 0.001 level with a Mann-Whitney-Wilcoxon test on RD and IQR pr). This shows that reasoning capabilities in Qwen3-14B causally impact the formation of speculative price bubbles. 

While Qwen3-14B without reasoning does not form speculative bubbles, it is still not well captured by the rational expectations hypothesis. Across the experimental runs, 90\% of all Qwen3-14B agents without reasoning are provably biased.

\subsubsection{Causal impact of non-linear extrapolation on bubble formation}

Our second causal question is about the causal effect of non-linear reasoning on bubble formation. We explore this by prompting models that do not form speculative bubbles to use non-linear methods. It should be noted that we therefore answer a slightly modified causal question; we isolate the causal impact of a prompt that asks the LLM to use non-linear methods, but this is slightly different from directly testing the causal impact of non-linear reasoning itself. The modified prompt may have other secondary impacts apart from increasing the amount of non-linear reasoning used.  The prompt used is below. 

\begin{promptbox}
Current time step: \{\}; below is a markdown table showing historical market prices and your predictions: 
                                    
\{\}

Total earnings up to time \{\}: \{\}. Earnings at last time step: \{\}.

Make your prediction for what the price will be in time period \{\}. Take your time, think step by step, and try to extrapolate observed trends into the future.  Pay particular attention to modelling second and higher order trends in the price. This could include quadratic models, exponential models, looking at the change in price differences or modelling the price returns rather than the absolute price.

Your response should be in the format:

\{"reasoning": "your reasoning here", "predictedValue": xx.xx\}
\end{promptbox}

We again focus on Qwen3-14B, this time using the non-linear prompt with Qwen3-14B without reasoning. With the regular, neutral prompt, Qwen3-14B without reasoning did not form speculative bubbles but with the non-linear prompt it sometimes does. Over 40 independent runs, Qwen3-14B with this non-linear prompt forms bubbles 52\% of the time. Further, the non-linear prompting causes higher overpricing (RD) and volatility (IQR pr) on average than regular, neutral prompting (significant at the 0.01 level with the Mann-Whitney-Wilcoxon test). While Qwen3-14B without reasoning and with non-linear prompting does not quite recover the Qwen3-14B with reasoning results, it does show that the non-linear prompt causally increases the propensity of Qwen3-14B without reasoning to form bubbles. 

\subsection{Analysis of agent prediction errors in single-LLM markets}
\label{app:results:agenterroranalysis}

We consider the mean squared error of agents’ predictions and decompose it into dispersion error (capturing heterogeneity across agents) and common error (capturing the deviation of the average prediction from the fundamental value) using Equation \ref{eq:agent_error}.

\begin{figure}[htbp]
    \centering
    \includegraphics[width=\linewidth]{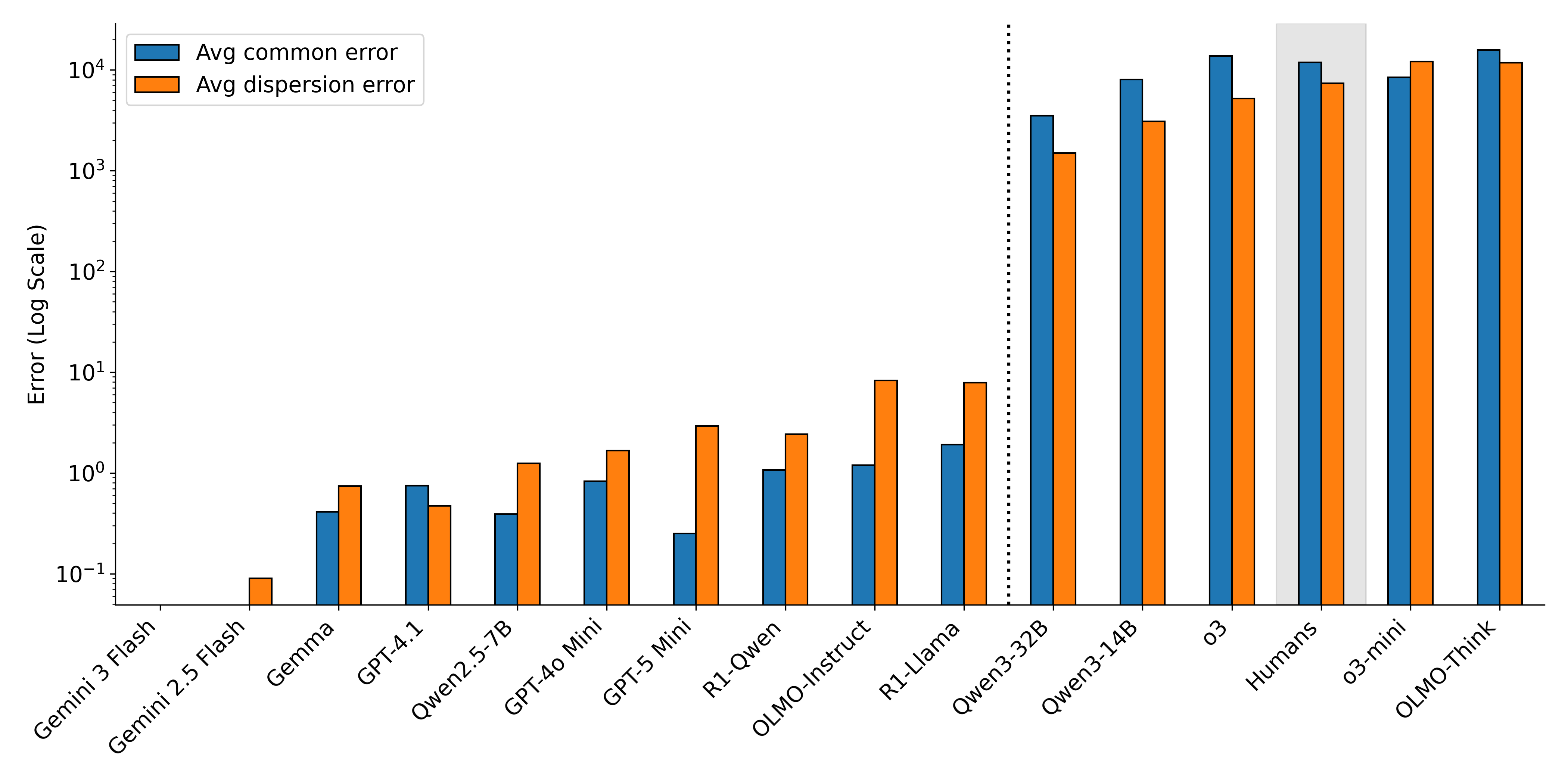}
    \caption{\textbf{Agents make varied predictions, although they tend to coordinate to an extent on (boundedly rational) strategies during bubbles.} Mean common and dispersion errors are plotted on a logarithmic scale for better visibility. Note that the bubble forming LLMs tend to have mean squared errors around 3-4 orders of magnitude bigger than LLMs that do not form bubbles. The dotted line divides models into those that do not form bubbles on the left, and those that do on the right.}
    \label{fig:agent_heterogeneity}
\end{figure}

We show the mean across experimental repeats of the dispersion and common errors in Figure \ref{fig:agent_heterogeneity}. The mean squared errors are 3-4 orders of magnitude larger for models that form bubbles than those that do not. This is expected as experiments with bubbles have far more volatility, making it harder to predict prices, and also have higher peak prices, where a small percentage error can translate into a large error in absolute terms.

Looking first at the five models that form bubbles, we see from Figure \ref{fig:agent_heterogeneity} that the common error tends to be larger than the dispersion error (except for o3-mini). This shows that agents coordinate on an incorrect and boundedly rational strategy. However, significant agent-level dispersion remains, showing the heterogeneity in prediction strategies used by different agents (of the same LLM). These results are similar to those from human experiments, highlighted in grey in Figure \ref{fig:agent_heterogeneity}.  Turning now to the LLMs which do not form bubbles, we see that the mean agent error is dominated by the dispersion error, showing once again the heterogeneity that even agents with the same underlying LLM can have.  However, there remains some common error, which demonstrates that even for models that do not form bubbles, agents coordinate on incorrect strategies. This means that even rational expectations with noise is not a good model for these LLMs. Agents coordinate on a boundedly rational strategy, even when they do not form large bubbles. The exceptions to this are both Geminis, and to some extent GPT-5 Mini. The Geminis have near zero common error, and GPT-5 Mini has the next lowest common error, reinforcing our findings in Section \ref{subsec:bounded_rationality} that these models exhibit behaviour (mostly) consistent with rational expectations.

\subsection{Classification of runs}
\label{app:results:mixedunpredictable}

We classify the experimental runs from mixed markets into 5 different categories to illustrate the breadth and variability of the market's macro behaviour. 

The first two categories are runs where no bubble is formed. We use the bubble definition introduced in this paper, which as a reminder, classifies an experimental run as containing a bubble if the price goes above 300 for 3 adjacent time steps or more. The first category, named \enquote{No bubble (low volatility)}, has no bubble formation and very low volatility throughout the 50 time steps. We quantify this as the standard deviation of the prices in the first 25 time steps and the last 25 time steps both being below 20. This category is generally markets where the fundamental price rational expectations hypothesis solution represents the experiment well. The second category, \enquote{No bubble (volatility)} is all other runs without a bubble, and generally represents markets where no bubble is formed but there is some volatility, most often when the rational expectations hypothesis is not a good model. 

The three remaining categories are all runs in which there is a bubble. We divide these into runs where only the first half of the time steps have a price standard deviation above 100, \enquote{Bubble (early volatility)}, only the second half of the time steps have a price standard deviation above 100, \enquote{Bubble (late volatility)}, or when both halves have a standard deviation above 100, \enquote{Bubble (persistent volatility)}. These generally represent a bubble forming early, late or consistent bubbles forming and reforming respectively. 

We note that these 5 categories are synthetic and used to illustrate how diverse the aggregate price dynamics are of this mixed market. In reality, even within these categories there are differences between experimental runs and we could include several other, more granular, categories.

\subsection{Mixed markets give varied results across temperatures}
\label{app:results:mixedmarketstemprobust}

We explore the same mixed market, with 6 different LLM agents, across the temperatures of 0.3, 0.7 and 1.0. We find that this mixed market continues to give varied behaviour, even with lower temperatures, showing that the variability of the macro outcomes of this market are a function of the heterogeneous agent prediction methods, rather than randomness from a high temperature. Using the method for classification of runs from Appendix \ref{app:results:mixedunpredictable}, we present the proportions of each category in the Table below:

\begin{table}[htbp]
    \centering
    \setlength{\tabcolsep}{12pt} 
    \renewcommand{\arraystretch}{1.2} 
    \begin{tabular}{l ccc}
        \toprule
        & \multicolumn{3}{c}{\textbf{Temperature}} \\
        \cmidrule(lr){2-4}
        \textbf{Metric} & \textbf{0.3} & \textbf{0.7} & \textbf{1.0} \\
        \midrule
        Repeats (Rpts) & 20 & 17 & 50 \\
        \midrule
        \rowcolor[gray]{0.95} \textit{No Bubble Group} & & & \\
        \quad Low volatility & 20\% & 35\% & 30\% \\
        \quad Volatility & 25\% & 24\% & 18\% \\
        \midrule
        \rowcolor[gray]{0.95} \textit{Bubble Group} & & & \\
        \quad Early volatility & 45\% & 12\% & 30\% \\
        \quad Late volatility & 0\% & 0\% & 12\% \\
        \quad Persistent volatility & 10\% & 29\% & 10\% \\
        \bottomrule
    \end{tabular}
    \caption{\textbf{Mixed market still gives varied results across temperature values.}  All experiments are with the same market, which contains 6 agents with different LLMs: Qwen3-14B, OLMO-Think, R1-Llama, Gemma, Gemini-3-Flash, and GPT-5 Mini. }
    \label{app:results:mixedmarketrobusttemp}
\end{table}

We find from Table \ref{app:results:mixedmarketstemprobust} that even when agents have very low temperature settings (0.3), they still form varied macro behaviour across experimental repeats. These results show that our findings are robust to the temperature parameter.

It is interesting to note that at lower temperatures there are no experimental repeats that form late volatility bubbles. This may be of interest to explore further in future work, as it may be that this particular type of behaviour requires extra randomness in the market to occur. 

\subsection{Data Leakage}
\label{app:results:dataleakage}

We first conduct a keyword search on all price prediction justifications across experimental runs, agents and time steps. Keywords which could indicate data leakage are listed in Table \ref{app:dataleakage:keywords}.

\begin{table}[ht]
\centering
\footnotesize
\begin{tabular}{lll}
\hline
Cars & Brock & prior experiments \\
Hommes & Heemeijer & previous experiments \\
adaptive learning & JEBO & well documented \\
learning to forecast & Journal & documented \\
learning-to-forecast & Velden & Expectations and bubbles in asset pricing experiments \\
LtFE & Rational Expectations & Sonnemans \\
LtF & experimental subjects & Tuinstra \\
human & laboratory & Journal of Economic Dynamics \\
humans & adaptive expectations & Journal of Economic Behavior \& Organization \\
prior results & previous results &  \\
\hline
\end{tabular}
\caption{\textbf{Keywords that would indicate data leakage.}}
\label{app:dataleakage:keywords}
\end{table}

o3 shows the most evidence of data leakage using the keyword search, with \enquote{Hommes}, \enquote{learning-to-forecast}, \enquote{human}, \enquote{laboratory}, \enquote{adaptive expectations}, \enquote{previous experiments} and \enquote{documented} all appearing. Most words only appear once or twice (out of 1800 justifications) but \enquote{laboratory} appears 311 times. However, \enquote{laboratory} was used by o3 in the context of trying to suggest behaviour to maximise earnings, rather than directly trying to re-create (or assuming) human behaviour. Meanwhile, Gemini-3-Flash was the only other model to use any of the listed keywords, mentioning \enquote{previous results} and \enquote{Rational Expectations} fewer than 11 times each, again out of 1800 justifications. 

We also test for data leakage by directly asking each model 5 times the below questions, followed each time by giving the model the experiment instructions. 

\begin{promptbox}
    Which experiment do you think the below instructions are most likely from?
\end{promptbox}

\begin{promptbox}
    The following instructions are from an experiment, what do you think is the main result of this experiment?
\end{promptbox}

\begin{promptbox}
    The following instructions are from an experiment in which price bubbles form. What do you think is the approximate maximum price of the bubbles?
\end{promptbox}

Frontier or near-frontier models such as o3, Gemini-3-Flash, GPT-5 Mini and GPT-4.1 all correctly identify the paper, authors and title of the experiment. These models, as well as Gemini-2.5-Flash correctly identify the key findings of the paper (although not for every repeat), giving answers such as \enquote{prices and predictions often diverge persistently from fundamental values} or \enquote{endogenous demand effects and boundedly rational forecasting behavior can produce bubbles}. However, none of the models correctly identify the bubble height as near 1000, generally guessing much closer to the 100-300 range. 

\end{document}